\documentclass[journal]{IEEEtran}
\usepackage{multirow}
\usepackage{cite}
\usepackage{graphicx}
\usepackage{psfrag}
\usepackage{amsfonts}
\usepackage{url}
\usepackage{stfloats}
\usepackage{float}
\usepackage{lipsum, color}
\usepackage{caption}
\usepackage{subcaption}
\usepackage{amsmath}
\usepackage[utf8]{inputenc}
\usepackage[table]{xcolor}
\usepackage{array}
\usepackage{xcolor}
\usepackage{latexsym}
\usepackage{color}
\usepackage{balance}
\usepackage{afterpage}
\usepackage{algpseudocode}
\usepackage{amsmath}
\usepackage{array}
\usepackage{tabularx}
\usepackage{multirow}
\usepackage{cite}
\usepackage{graphicx}
\usepackage{psfrag}
\usepackage{url}
\usepackage{stfloats}
\usepackage{amsmath}
\usepackage{xcolor}
\usepackage{soul}

\begin{document}
\title{Use of Federated Learning and Blockchain towards Securing Financial Services}  

\author{\IEEEauthorblockN{Pushpita Chatterjee\IEEEauthorrefmark{1}, Debashis Das\IEEEauthorrefmark{2},  and Danda B Rawat\IEEEauthorrefmark{3}\\
\IEEEauthorrefmark{1,3}{Department of EE and CS, Howard University, Washington, DC, USA \\
\IEEEauthorrefmark{2}{Department of CSE, University of Kalyani, Kalyani, India}\\
pushpita.c@ieee.org\IEEEauthorrefmark{1},
debashis.das@ieee.org\IEEEauthorrefmark{2},
db.rawat@ieee.org\IEEEauthorrefmark{3}
}
}
}
\maketitle 

\begin{abstract}
In recent days, the proliferation of several existing and new cyber-attacks pose an axiomatic threat to the stability of financial services. It is hard to predict the nature of attacks that can trigger a serious financial crisis. The unprecedented digital transformation to financial services has been accelerated during the COVID-19 pandemic and it is still ongoing. Attackers are taking advantage of this transformation and pose a new global threat to financial stability and integrity. Many large organizations are switching from centralized finance (CeFi) to decentralized finance (DeFi) because decentralized finance has many advantages. Blockchain can bring  big and far-reaching effects on the trustworthiness, safety, accessibility, cost-effectiveness, and openness of the financial sector. The present paper gives an in-depth look at how blockchain and federated learning (FL) are used in financial services. It starts with an overview of recent developments in both use cases. This paper explores and discusses existing financial service vulnerabilities, potential threats, and consequent risks. So, we explain the problems that can be fixed in financial services and how blockchain and FL could help solve them. These problems include data protection, storage optimization, and making more money in financial services. We looked at many blockchain-enabled FL methods and came up with some possible solutions that could be used in financial services to solve several challenges like cost-effectiveness, automation, and security control. Finally, we point out some future directions at the end of this study.
\end{abstract}


\section{Introduction}
\begin{figure}[!b]
\centering
\includegraphics[width=0.5\textwidth]{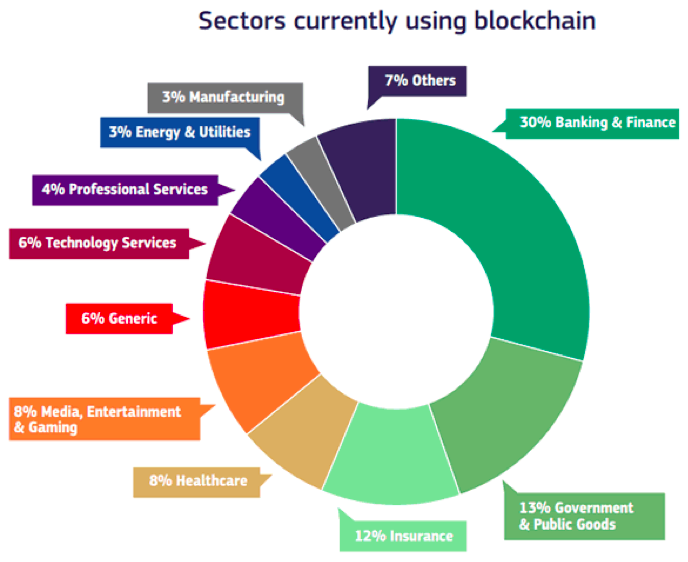}
 \caption{Usage of blockchain in financial sectors rather than other sectors. \cite{56}}
\label{fig1}
\vspace{-0.5cm} 
\end{figure}

Financial services have started to realize that blockchain technology could have a revolutionary effect on things like increasing revenue, improving the end-user experience and delivery process, increasing efficiency, and lowering the risks that come with running a business \cite{1}. The financial technology industry, much like every other sector focused on technology, is now in the process of developing. Fig. \ref{fig1} \cite{56} shows the use of blockchain in financial sectors is greater than in other sectors. There are a lot of new financial apps coming out all the time, and each one offers better and more creative ways to handle payments and process them. By the end of the year 2028, it is anticipated that the value of the financial blockchain industry will have increased to 36.04 billion dollars \cite{2}. Decentralized Finance, often known as ``DeFi," is an emerging financial technology that is based on blockchain and aims to limit the amount of control that banks have over financial services and money. Over many decades, digital ledgers will also cause big changes in how we get, send, store, and manage our money.

Almost any business may fall prey to cybercriminals, but they often choose their targets based on which ones would net them the most money or have the biggest impact \cite{4}. Hackers often go after banks and other similar financial services because they meet the above requirements. In the business of finance, companies often deal with billions of dollars and keep very sensitive and valuable information in the form of computerized archives like social security numbers, bank account details, wills, titles, and other personal documents. Their ongoing attempts to digitally change and the challenging policy landscape that is speeding up the use of hybrid workspaces all make it easier for cybercriminals to get data and sell it. As a direct consequence of this, cyber threat actors are focusing an inordinate amount of their attention on the banking industry.

\begin{figure}[!t]
\centering
\includegraphics[width=0.5\textwidth]{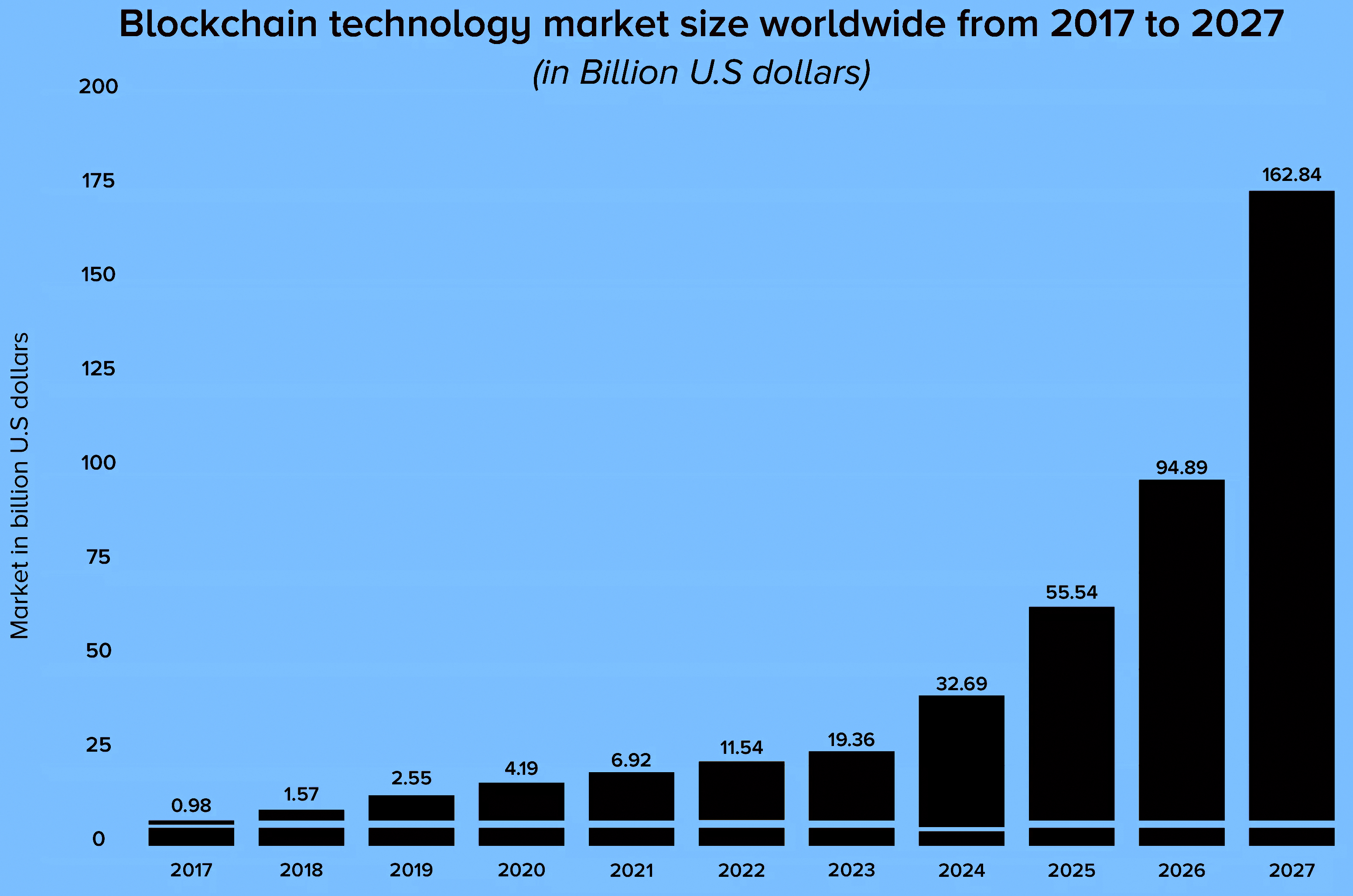}
 \caption{Expected blockchain usage in 10 years.  \cite{3}}
\label{fig2}
\vspace{-0.5cm} 
\end{figure}

The blockchain \cite{5} has a powerful function for storing certificates and can track data in a way that can be controlled. Federated Learning, on the other hand, can make sure that sensitive data never leaves the local node, that only the gradient information of the model update needs to be sent, and that all data is kept secret. Complex joint model training is done based on this premise, which makes these benefits possible. As a result, the functions that they perform are somewhat complementary to one another. If they can be combined, they will not only be able to share data more efficiently, but they will also be able to keep the privacy of the data. Blockchain usage has been increasing day by day, as shown in \ref{fig2}. Because blockchain \cite{6}is decentralized, immutable, and secure, combining these two technologies has the potential to improve Federated Learning's transparency, trustworthiness, and, most importantly, decentralization \cite{7}.

\par \noindent \textbf{Motivation}
The financial sector has benefited from several technological improvements and integrations. However, it still works in a centralized way, with financial institutions and governments at the center of the model. Customers of financial services have begun to question the relevance of this time-honored custom. Because of this uncertainty, blockchain development services have emerged as a more open and honest alternative. Because technology has added a new dimension, the financial services environment, which has become a tech-based revolution in the financial sector, now has a new facet. It has caused a lot of changes in how companies are set up and how they do business, which has given the financial technology sector a huge chance. Because of this, both new businesses and companies that have been around for a while and focus on making financial applications are interested in finding out if blockchain is necessary for financial services or not. According to Fig. \ref{fig2}, the use of blockchain in financial services is greater than in other sectors. A common factor that contributes to the complexity of risk is the fact that it is impossible to completely remove or guard against all risks, regardless of how sophisticated your systems are. It is where the process of risk management comes in. Risk management is a regular, ongoing process in which the right professionals look at risks from time to time to reduce the chances that certain threats will come true.
Companies that provide financial services in today's market not only have a hard time luring in new clients, but they also have a hard time doing the same with prospective workers. It may be hard to find the right people to fill new positions in information technology (IT) because of several factors, the most important of which is that millennials don't like long-term jobs. The provision of essential financial services is essential to the operation of any economy. Without them, those who have money to save may have difficulty locating others who need to borrow money, and vice versa. And without financial services, people might not buy as many goods or services because they would be so worried about saving money to protect themselves from possible losses.

\begin{figure}[!b]
\centering
\includegraphics[width=0.48\textwidth]{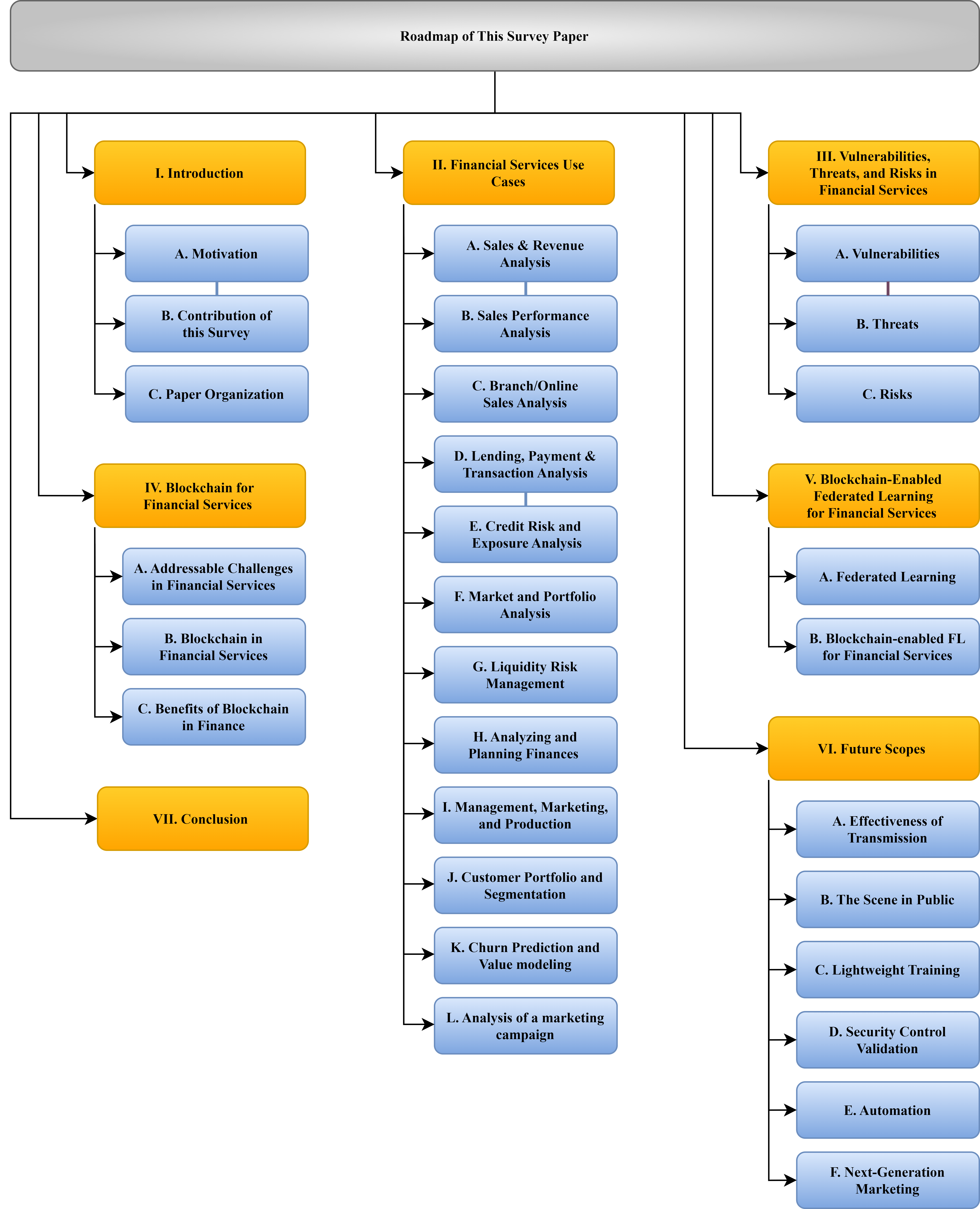}
 \caption{Roadmap of the survey.}
\label{fig3}
\vspace{-0.5cm} 
\end{figure}

\par \noindent \textbf{Contribution of the work}
This paper gives an overview of how blockchain and federated learning are used in the financial services industry. The goal of this research is to look at the benefits of blockchain and FL to get useful information that can be used to improve external statistics and make policy decisions in the financial sector. This paper talks about and analyzes how financial activities face weaknesses and problems. Besides, it is discussed how financial services can be used and organized using decentralized technology. However, the main contribution of this paper is presented as follows:

\begin{itemize}
    \item We give an overview of financial services by talking about how important they are and showing how they can be used in different situations.
\item We provide recent financial services vulnerabilities and related threats that have occurred. Possible risks in financial services are also discussed in this paper.
\item We demonstrate a few problems that exist in financial services and how blockchain can be used to solve them. The benefits of blockchain in this sector are also discussed in detail.
\item We present several blockchains and federated learning applications in different fields. Parallelly, we give some ideas for how blockchain and FL can be used in financial services and explain what these technologies signify.
\end{itemize}

The remaining of this paper is organized as follows. Fig. \ref{fig3} shows the overall roadmap of this survey paper. Section II gives the existing use cases of financial services. In section III, existing vulnerabilities, threats, and risks in financial services are provided in detail. Section IV depicts the usage of blockchain in financial services. Applications of Blockchain and federated learning in financial services are stated in section V. Section VI introduces some future research aspects in financial services. Finally, the paper is concluded in section VI.

\section{Financial Services Use Cases}
The Banking, Financial Services, and Insurance (BFSI) \cite{8} sector is the most vulnerable to uncertainty because it depends on global trends, changing laws, and customer demographics. It is a metric that has a great deal of significance for any sector, particularly the BFSI sector. Customers often give their banks and insurance companies their most private information, like information about their health or finances. It could have consequences for both the bank and the insurance company. Data and analytics are being used by the most important institutions in this field to change the rules of the game. They are gathering additional information from sources such as telecom providers, merchants, and social media to improve the knowledge that they already have about their customers. Because they have such a comprehensive perspective on their consumers, they are in a position to increase revenue, reduce risk, cut opportunity costs, and improve operational efficiency. 
\par \noindent \textbf{Sales \& Revenue Analysis}
Examining the operating procedures assists financial institutions in lowering their continuing expenses. If you know the sales trends for a certain consumer, you may be able to make things easier to repeat. Sales is a key activity, and having Business Intelligence (BI) tools may aid in defining benchmarks like the number of net new customers and the lucrative sector among current customers. These are just two examples of how having these tools can be beneficial.
\par \noindent \textbf{Sales Performance Analysis}
The report is all-encompassing and includes data on employee productivity. Any employee who works with customers, like salespeople, account managers, and tellers, can benefit from the information it can give because it can help them find ways to improve and, in the end, give better service to their customers. This evaluation could be used to check the viability of new financial products or services and make strategic changes that are in line with the institution's long-term goals.
\par \noindent \textbf{Branch/Online Sales Analysis}
It may assist financial institutions in formulating the most effective channel strategy. Multiple channels are now available for customers to use when communicating with their banks. Their trips through these channels are very complicated. They often start in one channel, go through different stages of the process in another channel, and then end up in a different channel. By collecting real-time data and using analytics to learn more about the buyer's journey, financial institutions may be able to use this to give customers a truly seamless multichannel experience. In addition to this, it assists them in maintaining an awareness of their rivals.
\par \noindent \textbf{Lending, Payment \& Transaction Analysis}
Banks can use their customers' transaction history to recommend products and services that are relevant to those customers. It leads to improved conversion rates as well as increased levels of client satisfaction. The following information about customers may be analyzed more effectively with the use of banking analytics: Existing clients of a bank can ask to look at their transaction history, which could include information about deposits, withdrawals, or payments. Bankers can help their customers take advantage of good deals on their credit or debit cards or other new financial products by getting to know how they spend their money and encouraging them to do so. Using this information to send timely spending alerts and payment reminders can improve the customer experience. It can be done to improve the customer experience. By looking at their clients' transaction histories and looking for patterns, banks can find transactions that might be fraudulent and take steps to stop them. Analytics for banking include data-driven methods like digital credit evaluation, improved early-warning systems, next-generation stress testing, and analytics for collecting debt. These techniques are used to protect clients against fraud at financial institutions.
\par \noindent \textbf{Credit Risk and Exposure Analysis}
An analysis of a client's credit risk and exposure might shed light on whether or not a customer has a history of defaulting on their payments in the past. These consumers' credit profiles highlight their assets and customer behavioral data, such as past-due bills, loans or borrowings, and earnings, that may be used to calculate each customer's credit score.
\par \noindent \textbf{Market and Portfolio Analysis}
It is highly important to do market and portfolio research to recruit new clients and keep the ones you already have. An analytics system could look at a client's current portfolio to suggest new investment options to the client and help the client's portfolio managers keep a steady return. Again, doing a market study is of the utmost importance when it comes to building a portfolio that will be successful regardless of the state of the economy.
\par \noindent \textbf{Liquidity Risk Management}
Every single banking procedure has the potential to become more efficient and streamlined. Using advanced analytics, financial institutions may be able to do things like answer questions from regulators faster and more accurately and give teams more information. It helps with decision-making that is augmented by analytics. The compliance and regulatory standards that banks must meet are quite severe. This has a big effect on how poor the impairment risk they face is. Know-your-customer (KYC) analysis is very vital, not only to achieve compliance with the legal requirements but also as a method of mitigating risk. Anti-money laundering (AML) analysts can more effectively detect and monitor problematic account holders with the use of these BI technologies.
\par \noindent \textbf{Analyzing and Planning Finances} 
Finance is the core of every company, just as it is with every other kind of organization. In the case of banks, this issue is even more important because bank employees are not only responsible for running the day-to-day business of the bank but also for meeting the different financial needs of customers. An analytical system may find the following use cases when it comes to a bank's finances: Banks need to have their cash on hand to handle payments well and follow all of the rules set by regulators. By looking at how much they spent in the past, they can make a budget that works for them. They also take into account certain factors that could make their financial needs go up or down. This could lead to finding a clear set of important success criteria that turn short-term savings into long-term, sustainable improvements and the best way to manage costs. Business intelligence (BI) tools could make financial planning and analysis (FP\&A) easier and make it easier to report to key stakeholders in a useful way. By making the necessary reports automatically and regularly, these systems could cut the amount of work needed for financial reporting by a large amount. In addition to this, they help speed up the transmission of information and ensure that decision-makers are kept up-to-date on the state of the bank's finances.
\par \noindent \textbf{Management, Marketing, and Production}
There are instances when a new strategy for approaching an established consumer is necessary. Banks need to give their current customers suggestions for new and better products, and this information should be given to these customers at the right time. When you look at the company's current customers, you can see which marketing methods have worked best in the past. You can then use these methods to bring in new customers. Business intelligence systems can be used with transactional and trade analytics to make more complete and richer profiles of customers. This, in turn, can increase the acquisition and retention of consumers as well as cross- and upselling opportunities.
\par \noindent \textbf{Customer Portfolio and Segmentation}
It is another significant use of analytics systems in financial services. It is necessary to correctly segment clients to successfully market to them. Consumers who are searching for a house loan or a vehicle loan are an example of one kind of customer segmentation used by financial services organizations. Another example would be customers who are specifically interested in a checking account or a money market account. Conversion can happen when the customer relationship manager makes an offer or calls the customer about something important to them. In the same way, a new offer may be aimed at a smaller group of people based on their credit scores.
\par \noindent \textbf{Churn Prediction and Value modeling}
Predicting a client's likelihood to churn and estimating their lifetime worth as a client are two areas that have gained major significance for financial institutions like banks and insurance companies in recent years. It takes a massive expenditure to compete with the thousands of businesses that are fighting for consumers' attention and physical space. It is of the utmost importance to make sure that you are not leaving any value on the table after you have successfully onboarded the consumer. The process of mapping the customer journey to observe their behavior helps in understanding any requirements that the customer may have and also assists in up-selling.

\par \noindent \textbf{Analysis of a marketing campaign}
An analysis of a marketing campaign gives a summary of the different channels that work for a bank and find the best way to spend money on all of them. The leading banks use the information from the transaction data of credit cards (from both their terminals and those of other banks) to develop offers that provide customers with an incentive to make regular purchases from one of the bank's merchants. These offers can be found on the websites of the leading banks.

\begin{figure}[!t]
\centering
\includegraphics[width=0.5\textwidth]{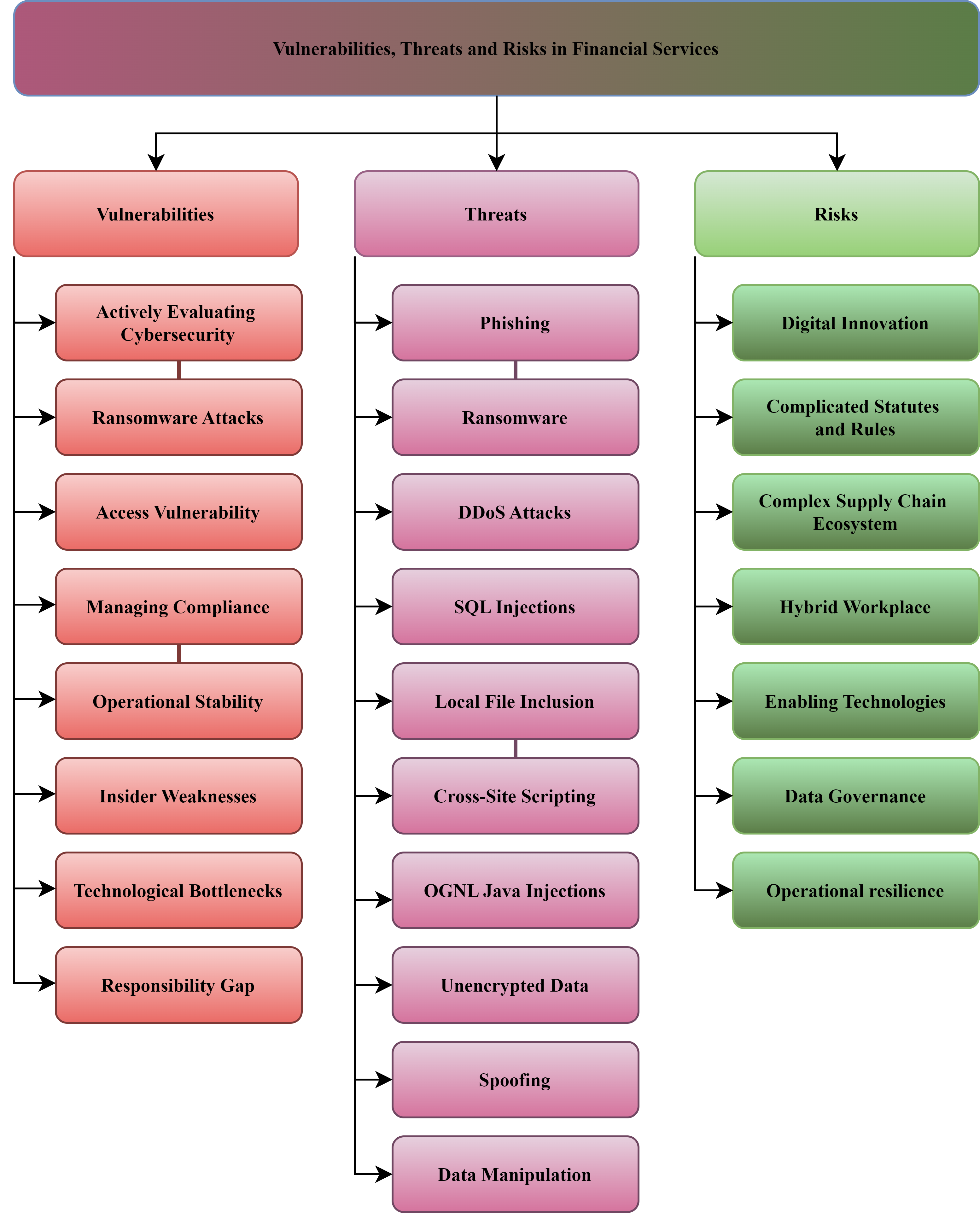}
 \caption{Vulnerabilities, threats, and risks in financial services}
\label{fig4}
\vspace{-0.5cm} 
\end{figure}

\section{Vulnerabilities, Threats, and Risks in Financial Services}
Financial services businesses produced more data after becoming digital. Every time you make a financial transaction or contact someone, data is created and shared using multiple applications. Cybercriminals can use this information to their advantage. They may sell, utilize, or threaten to dump it. In 2022, financial institutions are most at risk from ransomware, phishing, online application, and vulnerability exploitation attacks, denial of service (DoS) attacks, insider threats, nation-state and state-sponsored threat actors, and Advanced Persistent Threat (APT) groups \cite{9}. Fig.~\ref{fig4} shows financial services vulnerabilities, threats, and risks. Trends make financial services organizations more vulnerable to cyberattacks. Thus, firms must understand their current dangers and build effective defenses against them.
\subsection{Vulnerabilities}
Organizations in the financial industry confront security risks from both internal and external sources regularly because they are high-value targets for hackers. Threat actors use banking websites or virtual private networks (VPNs) to get into online banking systems to steal account information, cause trouble, or test how far they can get into a network. Internal threats often come from unhappy employees, weak third-party vendors, and human mistakes caused by phishing emails or other forms of social engineering. Weaknesses in external and internal security let sensitive financial information, client data, and monitoring networks for account balances and transactions get out. It hurts customer confidence and causes business problems\cite{9}. Today's financial services firms need sophisticated cybersecurity solutions that can manage the growing demands of keeping customer and financial data safe, limiting attack risk, and complying with regulatory regulations. Organizations in the financial industry may use Core Security's financial security services to identify and prioritize their most significant vulnerabilities and access concerns and effectively handle mandatory compliance audits.

\par \noindent \textbf{Actively Evaluating Cybersecurity}
Unknown weaknesses in cybersecurity and compliance cannot be addressed by financial organizations. In addition, failure to address these vulnerabilities may have significant repercussions. If left neglected until an event happens, institutions are compelled to use a reactive response, which may result in business interruptions and eroded client trust. Instead, financial service companies should take a proactive approach. Financial service organizations may conduct an initial evaluation of current vulnerabilities to discuss with a managed service provider (MSP).
\par \noindent \textbf{Ransomware Attacks}
There is an exponential increase in the number of potential targets for ransomware assaults as the globe continues to become more digitally linked. The term ``ransomware" refers to an attack in which the perpetrators employ malware to get access to your business's systems or data and then keep that data hostage until the firm pays a ransom. The aftermath of these assaults has been utterly catastrophic. In addition to the cost of the ransom, there may be additional expenditures related to damage management, such as legal fees and other expenses. There is also the possibility of losing data.
3)	Access Vulnerability
Sensitive data can be left exposed and subject to attack if there are flaws in the different levels of information access. Integration of cybersecurity measures is essential throughout all departments of a business and at each level of access. Criminals online will attempt to take advantage of whatever vulnerabilities they may find, regardless of the organizational hierarchy of the company they are targeting.
4)	Managing Compliance The advancement of information technology has made the financial services industry's job more difficult in terms of complying with regulations. The financial services industry in the United States is one of the most heavily regulated corporate sectors in the world. On the other hand, merely complying with the rules may not be enough anymore. Instead, aggressively managing compliance risk and increasing compliance overall is essential for gaining the trust of customers and avoiding expensive fines.
5)	Operational Stability
A backup and disaster recovery solution that is proactive and dynamic is necessary for avoiding disruptions to corporate operations and the loss of crucial data, either of which might result in a compliance violation. It is common for off-the-shelf onsite backup systems to be unable to provide the degree of performance necessary to satisfy the requirements of finance and investment businesses. It is essential to come up with a solution in advance of an outage to guarantee a speedy recovery and reduce the amount of time customers are without service.
\par \noindent \textbf{Insider Weaknesses}
Insider vulnerabilities are a source of concern in the banking and financial industries when it comes to cybersecurity. It occurs when people who work inside a bank or other financial institution do something that puts the company at risk of being attacked. The 2019 IBM X-Force Intelligence Index \cite{10} found that phishing emails were used in almost two-thirds (29\%) of the attacks that were looked at. Whaling attacks, which are often called ``corporate email compromise scams," were to blame for 45\% of these problems. In these incidents, hackers try to break into the email accounts of important people in the organization, like the CEO, to get the company to reveal private information. Another common thing that might happen is that systems and servers are set up incorrectly. 
\par \noindent \textbf{Technological Bottlenecks}
Websites and apps about banking and money make the architecture of the network as a whole more vulnerable. Researchers discovered that they were more likely to be hacked into banking and finance systems. Cross-site scripting (XSS) attacks \cite{10}, which allow attackers to run malicious code on a website or app, could happen to 80\% of the people who were tested. Then, the bad script could change the site's content by getting to the user's cookies and other sensitive information. Users are more likely to mistrust websites and programs that have vulnerabilities like these. So, so if businesses want to stay competitive, they should look into what steps they can take to protect their websites and apps.
\par \noindent \textbf{Responsibility Gap}
Even though the global financial system is becoming more dependent on digital infrastructure, it is not clear who is responsible for protecting it from cyberattacks. It is due, in part, to the rapid pace at which the environment is changing. If people don't work together, the global financial system will continue to get worse as more innovation, more competition, and the pandemic speed up the pace of the digital revolution and make it more dangerous. Although many threat actors are motivated by a desire to make money, the number of attacks that are solely disruptive and destructive has been increasing; additionally, those who learn how to steal learn about the financial system's networks and operations, allowing them to release more obstructive or dangerous future attacks \cite{11}. Even though the system is generally well-developed and well-regulated, this sudden change, like the risks it poses, puts a strain on its ability to respond.

\subsection{Threats}
The worst catastrophes have harmed financial information, particularly accounts, calculations, and transactions. Such assaults, which may undermine trust, now have some technological solutions. VMware reported 238\% more financial institution cyberattacks in the first half of 2020 \cite{12}. IBM and the Ponemon Institute estimate a financial data breach would cost \$5.72 million in 2021 \cite{12}. It's global. The increased frequency of assaults on targets of opportunity in low- and lower-middle-income countries is less reported than cyberattacks in high-income countries. Financial inclusion has been the biggest driver of digital banking services like mobile payment systems. Digital banking services expand financial inclusiveness but provide hackers with more targets.

\par \noindent \textbf{Phishing}
Phishing is social engineering that deceives individuals into sharing their login credentials to enter a private network. Email phishing, when victims get official-looking emails, is the most common. Visiting a phishing email's dangerous links or attachments might install malware or launch a bogus website that steals login credentials. In the first half of 2021, bank phishing attacks rose 22\%. Financial app assaults increased by 38\% at the same time. Akamai's 2019 State of the Internet report found over 50\% of phishing attempts targeted financial services \cite{12}. Phishing tactics are evolving to exploit modern worries. These troubling trends rank phishing among the banking sector's top cybersecurity dangers.
\par \noindent \textbf{Ransomware}
Ransomware also threatens financial institutions. Ransomware encrypts computers, locking victims out \cite{12}. Only a ransom can fix the harm. Due to strict rules requiring financial institutions to be resilient to cyberattacks and data breaches, these extortion methods work effectively against them. Ransomware attacks are now data breaches, which might affect regulatory compliance requirements. Ransomware gangs target financial businesses because of their customer data. Due to the danger of data exposure on the dark web and reputational damage, many financial services companies accept extortion demands..
\par \noindent \textbf{DDoS Attacks}
2020 witnessed the highest DDoS attacks on financial institutions. DDoS attacks are a prevalent cyber threat to financial services since they may target consumer accounts, payment gateways, and banks' IT systems. Due to this, the impact of DDoS attacks on financial firms is amplified. Cybercriminals may use the ensuing confusion in one of two ways. Password login attacks and DoS attacks were the two main online dangers to payment systems in 2020. In comparison to the same period in 2020, multi-vector DDoS attacks have increased by 80\% in 2021 \cite{12}. These DDoS attacks combine several campaigns to swamp security personnel.
\par \noindent \textbf{SQL Injections}
A vulnerability in a WordPress plugin that enabled Time-Based Blind SQL injections (SQLi) was found in March of 2021 \cite{13}. This vulnerability was detected. There was a possibility that 600,000 customers were affected by this issue. Through the use of a technique known as Time-Based Blind SQLi, the vulnerability made it possible for any site visitor to access sensitive data stored in a website's database. Because the SQL query was executed inside the function object for the pages" page, this meant that any site visitor, even those who did not have a login, may trigger the execution of this SQL query. It would thus be possible for a hostile actor to give harmful values for either the ID or type parameters.
\par \noindent \textbf{Local File Inclusion}
A vulnerability known as Local File Inclusion (LFI) was discovered in August 2021 for a version of BIQS \cite{12} software used by driving schools for billing customers. When a certain payload is sent to download/index.php in older versions of BIQS IT Biqs-drive than v1.83, a local file inclusion (LFI) vulnerability is present. This vulnerability may be exploited to take control of the affected system. Because of this, the attacker can access arbitrary files stored on the server using the permissions of the web-user configuration.
\par \noindent \textbf{Cross-Site Scripting}
Trend Micro revealed the details of e-commerce website cross-site scripting (XSS) attacks on April 28, 2021. EC-CUBE-built websites have also had XSS instances confirmed by JPCERT/CC (an open-source CMS for e-commerce websites). Any e-commerce website having an XSS vulnerability on its administrator page is targeted by this attack. This attack campaign continues on July 1, 2021. In order forms on targeted e-commerce websites, attackers insert malicious scripts to make purchases. XSS attacks on the administrator's page steal credentials and install Simple WebShell on the website. Attackers then utilize WebShell and JavaScript on the website to harvest and save user data. Monitoring the WebShell may allow the attackers to obtain the stolen data. During the attack, the attackers embed Adminer \cite{15} on the e-commerce website. GUI-based database content analysis tool. It supports MySQL, PostgreSQL, SQLite, MS SQL, Oracle, SimpleDB, Elasticsearch, and MongoDB. Attackers presumably accessed database information using this approach.
\par \noindent \textbf{OGNL Java Injections}
In August 2021, OGNL flaws allowed hostile actors to inject code into Atlassian Confluence servers \cite{16}. OGNL injection vulnerabilities allow unauthenticated users to execute arbitrary code on Confluence Server or Data Center instances. Previous versions of the Confluence Server and Data Center are affected by this problem. The vulnerability is actively abused in nature. Unauthenticated users may exploit it regardless of settings.
\par \noindent \textbf{Unencrypted Data}
When data is left unencrypted \cite{17}, fraudsters or hackers may immediately change it, causing major problems for banks. Online and financial institution data must be jumbled. It prevents attackers from using stolen data.
\par \noindent \textbf{Spoofing}
It is one of the most recent instances of a cyber threat that businesses in the financial sector need to be prepared for. The URL of a bank's website will be impersonated by hackers, who will replace it with a website that is connected to the actual one and operates in the same manner (cite 17). When a customer uses a fraudulent website and inputs his login information, the hackers will grab the customer's credentials and utilize them in the future.
\par \noindent \textbf{Data Manipulation}
One of the most common misunderstandings about cyber assaults is the belief that people are only concerned about the theft of data. It isn't always the case, though, because hackers are using data manipulation attacks more and more. Cybercriminals are always developing new methods of attack. Attacks involving data manipulation happen when a bad actor gains access to a trusted system and then makes changes to the data without being caught to help themselves \cite{17}. One example of this would be if an employee changed information about customers. Likely, it won't be found out because the transactions will look like they were done legally. It will cause future data to be stored incorrectly. The more time that goes by before the manipulation is discovered, the more damage it will do.

\subsection{Risks}
The financial sector is getting more and more exposed to ``cyber risk," which is the risk of losing money because of how much they depend on computers and digital technology. Cyber-related events, especially cyberattacks, are always at the top of polls that measure the financial stability of the United States and the rest of the world. Cyber risk, like other financial vulnerabilities, raises macroprudential issues. Similar to other financial problems, a lot of technological attention has been paid to cyber resilience, but it is still very early to measure the effects that cyber risk might have on the financial system. If you want to be strong against cyber-attacks, you need to know about the problems that make the cyber risks the financial sector faces even higher. It is important to find a way to solve them all at once, as these problems are linked to each other.
\par \noindent \textbf{Digital Innovation}
In financial institutions (FIs), new technologies are being used, such as cloud computing, artificial intelligence, and digital service delivery. Most FIs are improving their data processing, fraud detection, and financial analytics by using software that is hosted in the cloud. Meanwhile, the COVID-19 epidemic furthered the process of transferring the industry's IT infrastructure (digital transformation), which resulted in the proliferation of virtual banks and financial services. Because of digital transformation, businesses today run an increasing number of brand-new apps, devices, and infrastructure components, all of which expand the attack surface. A surge in cybersecurity threats for financial institutions is caused by all of the issues together. Even if the rise of new technologies in the financial sector has a major impact on industrial risk management, these technologies could help risk management by improving cybersecurity and compliance controls.
\par \noindent \textbf{Complicated Statutes and Rules}
As financial institutions use more technology and data to help their customers, regulations must change to keep up. State, federal, and international authorities have established several new restrictions for their industries in reaction to the growth in cyberattacks on financial services organizations. Data protection, privacy, and cybersecurity legislation for financial institutions (FIs) is tightening in various nations. Compliance may be time-consuming and expensive, but it's in everyone's best interest. According to BITS' technology division, chief information security officers spend 40\% of their time addressing regulatory agency criteria \cite{18}. Because of the regulatory environment's complexity, enforcement is tighter, raising regulatory costs and penalties. In August 2020, the US government fined Capital One \$80 million for failing to find and deal with cyber risk, which led to a massive data breach in 2019 \cite{19}. Capital One resolved a class-action lawsuit in late December 2021 over a 2019 Amazon Web Services cloud network intrusion that stole 100 million customers' data \cite{20}. The settlement was for 190 million dollars.
\par \noindent \textbf{Complex Supply Chain Ecosystem} Most financial firms outsource their digital duties. Third-party service providers may be vulnerable even if the FI's internal security is strong. Threat actors are targeting software businesses and sending malware to supply chain customers through legitimate downloads and upgrades. Threat actors gained backdoor access to client networks via these attacks on software distribution platforms. Recent assaults include the SolarWinds breach \cite{21}. Supply-chain assault. Attackers infiltrated SolarWinds' network and planted malware in their management software to target thousands of banks and government entities. The SolarWinds breach shows how susceptible the financial services sector is to cyberattacks and disruptions since it depends on third-party suppliers and service providers with little or no cybersecurity oversight. Third-party cybersecurity vulnerabilities will grow as the government prioritizes business continuity and operational resilience.
\par \noindent \textbf{Hybrid Workplace} COVID-19 has sped up recent changes in the way people work, like the hybrid workspace, which combines people who work in the office and those who work from home. It will increase the risk that businesses face. As we move into the third year of the pandemic, more and more people are using technologies like remote work, hybrid workforces, and software that is hosted in the cloud. Businesses had no choice but to quickly adopt the new technologies that gave them remote access, better communication, and more ways to work together. Because of this, hybrid working settings make IT systems more complicated, increase the number of ways to attack them, and create new cyber risks and threats.
\par \noindent \textbf{Enabling Technologies} According to some estimates, the pandemic sped up the transition to digital technology by as much as three years. Enabling technologies, such as application programming interfaces, big data analytics, artificial intelligence, biometrics, cloud computing (particularly outsourcing to the cloud), and distributed ledger (blockchain) technology, makes it feasible for digital transformation to occur. Companies and their boards of directors need to be able to make sure that new technologies are adopted safely so that the benefits can be gained and the risks that come with trying new things can be managed proactively. This will help businesses get the most out of their innovative activities and reduce the risks that come with them.
\par \noindent \textbf{Data Governance}
The importance of having a solid strategy for data governance is only going to grow in the coming years. Companies need to realize that data is a key strategic asset before they can come up with a company-wide plan for collecting, managing, storing, protecting, retrieving, and destroying data. To put it another way, develop a strategy for data governance that is tailored to your organization. If data governance works, it will have many benefits, such as making it easier to see risks in a hybrid work environment, being able to meet the recently agreed-upon requirements for reporting climate risk, and making it easier to keep track of records.
\par \noindent \textbf{Operational Resilience}
Cybersecurity is a major problem for businesses operating in the financial industry. In a September 2021 Conference of State Bank Supervisors (CSBS) study, more than 80\% of bankers regarded cybersecurity risk as "very significant" as the top internal risk \cite{22}. This number is more than twice any other operational risk category and greater than the 60\% recorded last year. This risk aversion may be attributed to a great number of different factors. For example, worries about cybersecurity can hurt both the way a company works and its reputation. If a financial institution is hit by a cyberattack, its ability to do business could be hurt or completely stopped. It is called operational risk. In addition, as a result of the hack, consumers can lose faith in the company and want to conduct their business elsewhere (reputational risk).

\begin{figure}[!t]
\centering
\includegraphics[width=0.5\textwidth]{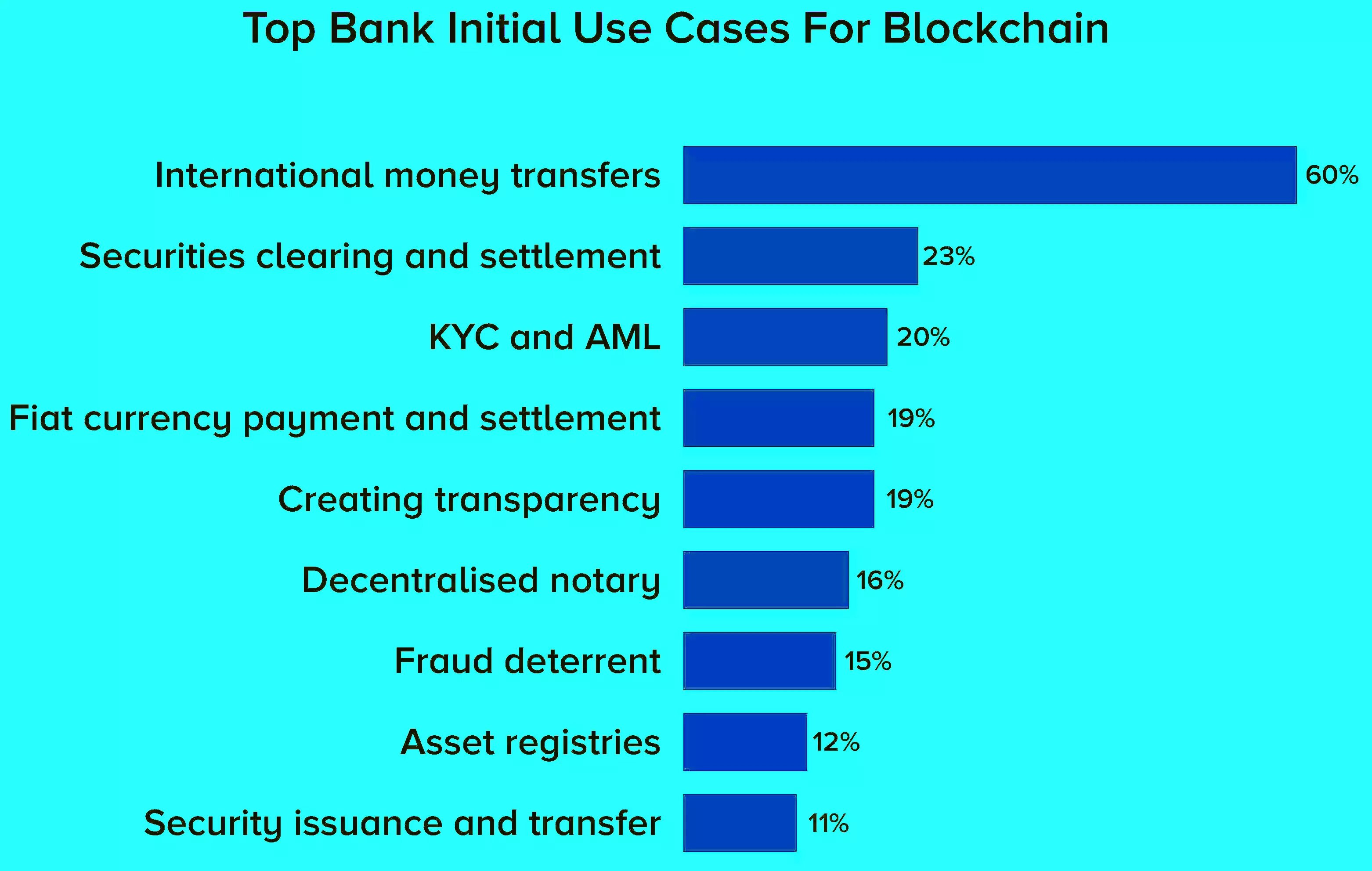}
 \caption{Usage of blockchain in different scenarios of financial services. \cite{3}}
\label{fig5}
\vspace{-0.5cm} 
\end{figure}

\section{Blockchain for financial services and banking industries}
Blockchain can make banking and lending easier by reducing the risk of the other party and the time it takes to issue and settle the money. Authenticated paperwork and KYC/AML data reduce operational risks and enable real-time financial document verification. Without a bank or financial services provider, a blockchain lets people transmit money securely \cite{57,59}. Blockchain technology is known as ''distributed ledger technology" in the financial services business \cite{58}. Table \ref{table:1} shows the existing methods using blockchain in financial services. Since all transactions on a blockchain are saved in a shared database, it could make banking more open. Because of this openness, problems like fraud might be found and fixed, which could make the risk for financial institutions lower \cite{60}. Fig. \ref{fig5} shows the usage of blockchain in several cases in financial sectors.

\begin{table*}[!hbp]
\scriptsize
\begin{center}
\caption{\textbf{Existing blockchain-based methods for financial services}}\label{table:1}
\begin{tabular}{ |m{3em}|m{13.5em}|m{15.3em}|m{16.3em} |m{16.3em}|} 
\hline
\textbf{Author / Ref} &\textbf{Proposed Method} &  \textbf{Description} &  \textbf{Pros} & \textbf{Cons} \\
\hline 

{Liu \textit{et al.,}\newline \cite{43}} & {Hybrid chain model combining PANDA and X-alliance}
 & {Hybrid chain model may handle each account's transaction in parallel, asynchronized from other adjacent accounts in the network. It provides efficient data storage and authorization control, and ownership of change tracking data} &{$\bullet$ High performance.
\newline $\bullet$ Lower protection cost} &{$\bullet$Performance accuracy is low. \newline $\bullet$ Increased sample data. \newline $\bullet$ Smart contract adoption issue.}
 \\
\hline

{Lorenz \textit{et al.,}\newline \cite{44}} & {Money laundering detection}
 & {Active learning approach that matches a fully supervised baseline with 5\% of labels. } 
 &{$\bullet$ Detect money laundering.
\newline $\bullet$ Simulates a real-world situation with few analysts for manual labeling.} 
&{$\bullet$ Lack of proper Anti-Money Laundering (AML) processing in all financial services.}
 \\
\hline

{Chen \textit{et al.,}\newline \cite{45}} & {Blockchain-enabled Financial
Surveillance Systems} & {Research on blockchain credit information preservation and supervision, post-loan management, and time-based financial supervision chain.} 
 &{$\bullet$ Safeguarding financial and user data.
\newline $\bullet$ Cost reduction in auditing.
\newline $\bullet$ Provides repayment ability function.} 
&{$\bullet$ User identity verification.
\newline $\bullet$ Unauthorized person can read financial data}
\\
\hline

{Chen \textit{et al.,}\newline \cite{46}} & {Online P2P lending}  & {Used machine learning and neural networks to forecast online lending credit risk.} 
 &{$\bullet$Improve the efficiency of online lending.
\newline $\bullet$reduced credit risks.
\newline $\bullet$increased trust in the models
} 
 &{$\bullet$ Time consuming.
 \newline $\bullet$ Prediction accuracy is lower.
}
\\
\hline

{Kher- \newline bouche \textit{et al.,}\newline \cite{47}} & {Combination of UML-AD and Blockchain}  & {Developed a blockchain UML profile, activity diagram, and Petri net verification to ensure a distributed computing system with modifiable information.} 
 &{$\bullet$Integrity maintenance for financial organizations. 
\newline $\bullet$Facilitate insurance claiming.
} 
 &{$\bullet$ Lack of financial business policies.
\newline $\bullet$ Patterns collection is minimal.
}
\\
\hline

{Ma \textit{et al.,}\newline \cite{48}} & {Blockchain-based Supply chain finance}  & {Hyperledger Fabric privacy security and supply chain financing.} 
 &{$\bullet$ Designed privacy protection mechanism.
} 
 &{$\bullet$ Lack of data privacy.
}
\\
\hline

{Du \textit{et al.,}\newline \cite{49}} & {Blockchain-enabled Supply chain financing}  & { The suggested supply chain finance platform overcomes the issue of non-trust among supply chain players and uses homomorphic encryption on the blockchain to safeguard confidential information in financial services.} 
 &{$\bullet$ Increase capital flow and data flow efficiency.
\newline $\bullet$ Protect user privacy and data privacy.

} 
 &{$\bullet$ Inadequate for massive data processing. \newline $\bullet$ Higher computational cost.
}
\\
\hline

{Wu and Duan \newline \cite{50}} & {Blocckhain for small and micro enterprises}  & {proposed that Xi'an's tiny and micro firms employ blockchain technology to solve their funding problems.} 
 &{$\bullet$ Enhance the asymmetry of information.
\newline $\bullet$ Additional financing assistance for small and medium-sized enterprises.
} 
 &{$\bullet$ Weak incentive policies. \newline $\bullet$ Privacy breach.
}
\\
\hline

{Camino \textit{et al.,}\newline \cite{51}} & {Financial Transaction Suspicions}  & { A methodology for doing financial data analysis that does not involve the use of ground truth.} 
 &{ $\bullet$ Detection of expected anomalies. \newline $\bullet$ Exploratory data analytics for financial data sources.
} 
 &{$\bullet$ Vast and challenging to comprehend.
}
\\
\hline

{Lyu \textit{et al.,}\newline \cite{52}} & {Blockchain-enabled regulatory sandbox}  & {Proposed a "regulatory sandbox" and industry norms immediately. } 
 &{$\bullet$ Improve banking efficiency.
\newline $\bullet$ Increase security in banking industries.
\newline $\bullet$ Solve regulatory problems.
} 
 &{$\bullet$ Lack of authenticity.
\newline $\bullet$ Lack of reliability.
}
\\
\hline
{Chod \textit{et al.,}\newline \cite{53}} & {b\_verify - an open-source software protocol}  & {Blockchain technology's transparency into a firm's supply chain allows it to get attractive financing conditions at reduced signaling costs. } 
 &{$\bullet$ security of financing conditions.
\newline $\bullet$ Less signaling cost.
} 
 &{$\bullet$ It is complex for humans to read cryptographic verification proof.
}
\\
\hline
{Li \textit{et al.,}\newline \cite{54}} & {Fabric-SCF}  & {A Blockchain-based secure storage solution using distributed consensus to protect, trace, and immutably store supply chain financial information.} 
 &{$\bullet$ Provides dynamic, fine-grained access control.
\newline $\bullet$ Secure storage management.
} 
 &{$\bullet$ Complex business logic is required instead of simple SCF logic. \newline $\bullet$ Throughput verification is required.
}
\\
\hline
{Li \textit{et al.,}\newline \cite{55}} & {Fabric-Chain \& Chain}  & {Proposed a blockchain and bloom-based storage mechanism for the electronic document data utilised in the supply chain finance business process.} 
 &{$\bullet$ Improved access efficiency.
\newline $\bullet$ Increased retrieval efficiency.
} 
 &{$\bullet$ Lack of portable business logic. \newline $\bullet$ lack of data updating.
}
\\
\hline
\end{tabular}
\end{center}
\end{table*}

\subsection{Addressable Challenges in Financial Services}
Financial services often have problems, like not reaching their goals, taking a long time to raise money, and losing more and more money. These problems are often caused by inadequate management. The following is a list of challenges that blockchain technology has the potential to solve in the financial technology industry \cite{23, 74}:
\par \noindent \textbf{Centralized System}
Even though financial services solutions made things seem easier, the real power was still in the hands of third parties \cite{65,66}. Higher-ups are still the only ones who can approve transactions, so users are still waiting for confirmation that they can move forward with their transactions. Because of the introduction of blockchain technology, this was the first problem in the financial services industry that could be addressed.
\par \noindent \textbf{Trust Issues} When consumers take any action inside financial services apps, they are not aware of what is occurring on the other side of the transaction \cite{67}. This leads to a great deal of uncertainty as well as an increase in the fear of having one's identity stolen, which eventually results in a decrease in faith in the process. Because the blockchain is open and can't be changed, these blockchain application development services solve this problem in the field of financial technology.
\par \noindent \textbf{Less Efficient Methods} One further reason why the financial technology industry requires blockchain is that the presence of many different third parties often causes the procedures to be delayed. This, in the end, leads to poorer rates of customer satisfaction and increased levels of volatility in the commercial sector as it is generated a lot of data \cite{64}.
\par \noindent \textbf{Higher Operational Cost} In the financial technology industry, time is money. As a result, blockchain technology has once again shown itself to be one of the financial services innovations that have the potential to lower costs by almost half. It is because it reduces the dependence on many individuals, makes the process public, and shortens the required time.
\subsection{Blockchain in Financial Services}
When talking about how blockchain technology has changed the financial technology industry, it's best to focus on the most important parts of the economy to better understand and analyze the changes. The following is a list of blockchain use cases for financial services \cite{24}.
\par \noindent \textbf{P2P Payments} Bank clearing and settlement regions have concerns about costly bureaucracy and unclear expertise. These concerns are present in most financial arrangements and cause concern. Old and hierarchical financial systems produce these gaps. Decentralized consensus methods can close them quicker. Blockchain technology helps financial services. Decentralized ledger technology will enable mobile banking for those without bank accounts. A blockchain mobile app development business may simplify cross-checking data between companies engaged in international payment transactions. Blockchain technology allows several checks at once.
\par \noindent \textbf{Financial Trading}
Documents are still being sent or faxed to confirm information that is necessary for trade financing, which means that paperwork is being sent across the world to verify the information. To buy stocks or shares, you still have to go through the complicated and time-consuming steps of brokerage, exchanges, clearing, and settlement. The settlement process takes three days on average, but it may take longer if it occurs over the weekend. This is because every trader is required to keep databases for all of the transaction-based documents, and they must routinely check these databases against each other to ensure that they are accurate. The application of blockchain technology to the provision of financial services in this sector has the potential to free traders from the need to do time-consuming checks on counter-parties while also improving the efficiency of the whole lifecycle \cite{62, 64}. This not only speeds up the settlement process but also makes transactions more accurate and reduces the risks involved \cite{63}.
\par \noindent \textbf{Crypto Lending} Thanks to crypto lending, the financial world now has a new, easy, and transparent way to lend money. The lenders will give the borrowers the assets they need for the loan at a rate of interest that was agreed upon ahead of time. Borrowers will be able to keep their crypto assets as collateral for a loan based on fiat currency or stablecoins. It is also true when read backward. When borrowers need to borrow crypto assets, they will occasionally use their stable coins or traditional cash as collateral.
\par \noindent \textbf{Regulatory Compliance} For the second time, this is one of the most consequential uses of blockchain in the financial sector. Since it is expected that the global need for regulatory services will expand in the next few years, financial services businesses are integrating blockchain technology to improve regulatory compliance \cite{68}. They expect this technology to record the actions of all parties involved in every verified transaction, eliminating the need for regulators to verify the records' veracity. Technology is also allowing scholars to return to the original documents rather than relying on the many copies that have been produced. Errors are less likely to occur, the integrity of records for financial reporting and audits is being preserved, and the time and resources spent on auditing and accounting are being drastically reduced thanks to the blockchain's promise of immutability \cite{69, 70}.
\par \noindent \textbf{Digital Identity} The number of accounts that have been made with fake information keeps going up. Even though banks do have stringent Know Your Customer and Anti-Money Laundering inspections, these measures are not failsafe. Customers are much less likely to be hacked because there isn't a clear set of documents, they need to provide to prove who they are. A digital identification system may benefit from using blockchain technology. The customers only need to go through the validation process once, and then they may use their credentials to conduct transactions in any part of the world. On this front, blockchain may also aid financial users in the following ways: 1) managing personal information; 2) communicating personal information to other parties while minimizing security concerns; and 3) digitally signing legal documents, such as claims and transactions \cite{75,76}.
\par \noindent \textbf{Auditing}
It is a procedure that checks the finances and brings to light any discrepancies that may exist. The procedure is not only difficult to understand, but it also moves quite slowly. Blockchain technology, on the other hand, makes the procedure simpler. Because of this technology, you can ask the blockchain application development firm with whom you are paired to add the record straight to the ledger, making it possible to see and update data in a time-efficient manner \cite{71}.
\par \noindent \textbf{New Crowdfunding Models}
The concept of "crowdfunding" refers to a method of supporting a project by soliciting contributions from a large number of individuals, often via the internet. ICOs, IEOs, and other mechanisms may make the fundraising process using blockchain technology more open and efficient, as opposed to more traditional methods of financing. However, it is highly recommended to have a clear understanding of what all financial services organizations that use blockchain technology are doing with it.

\subsection{Benefits of Blockchain in Finance}
Blockchain technology has made it possible to create inclusive, open, and safe corporate networks. These networks make it possible to issue digital security credentials in a shorter amount of time, at lower unit prices, and with a higher degree of customization \cite{72, 73}. Over the past few years, the use of blockchain technology in the financial sector has grown, which has shown the following benefits \cite{25,26,27}.
\par \noindent \textbf{Transparency}
Protocols, standards that everyone agrees on, and common procedures are all used in blockchain technology. Together, these things serve as a single source of growth for all members of the network. It makes the data more reliable and improves the customer experience by making processing faster.
\par \noindent \textbf{Security} In the financial sector, blockchain technology has made it possible to use secure application code that is designed to be impossible to change by hostile or third-party actors \cite{61}. It makes it practically impossible to modify or hack the system.
\par \noindent \textbf{Trust} The immutable and clear ledger makes it easier for the different people in a business network to work together on data management and get along. The blockchain is a distributed ledger technology that enables the safe recording, management, storage, and transmission of transactions across a wide variety of industries.
\par \noindent \textbf{Privacy} When blockchain technology is used in the financial industry, it protects data privacy at all levels of software stacks in a way that is unmatched in the industry. This makes it possible for businesses to share data selectively within their networks. This increases confidence and openness while preserving anonymity and privacy at the same time.
\par \noindent \textbf{Programmability} It makes it possible to design and run smart contracts, which are pieces of software that automate business logic and can't be changed. This makes them easier to program, more efficient, and more trustworthy.
\par \noindent \textbf{Scalability and High Performance}
In the financial sector, blockchain technology is used through hybrid and private networks that were built to handle hundreds of transactions per second. It provides enterprises with significant resilience and worldwide reaches by completely supporting interoperability across the public and private sectors.

\section{Blockchain-Enabled Federated Learning for Financial Services}

With a central server, FL has reached a point where it can't do anymore. At the same time, new risks are showing up. There are a lot of things going on, but the most important ones are centralized processing, making up data, and not having any incentives. Businesses and academics are paying a lot of attention to blockchain-enabled distributed ledger technology in the hopes that it will speed up the use of FL. There have been a lot of creative answers made to meet the needs of a wide range of situations that are always getting more complicated. The functionality of FL that is allowed by blockchain offers both ideas and approaches to increasing the functionality of FL from a variety of points of view.

\subsection{Federated Learning}
Federated learning (FL) is an exciting new decentralized deep learning technique that lets users update their models together without having to share their data. FL is changing how mathematical modeling and analysis are done in the business world. This makes it possible for more and more industries to build distributed machine learning models that protect the privacy and are safe. Nonetheless, the properties that are intrinsic to FL have resulted in several issues, including those about the security of personal information, the expense of communication, the heterogeneity of the systems, and the unreliability of model upload during real-world operation. It's interesting to think about how adding blockchain technology could improve FL's security and performance and increase the number of things that can be done with it. FL is a strategy for training artificial intelligence systems using data that is kept confidential. It lets centralized AI systems learn from data, which is often personal, without the data's actual content being shared or made public. Instead, only the lessons that can be gleaned from the data's structure are used.

\subsection{Blockchain-enabled FL for Financial Services}
The conventional architecture of federated learning is based on centralization. But a model that depends on a trustworthy central server has several security holes. If the central server is attacked, the whole federated learning process could be ruined. This is because the global model update wouldn't be able to happen, training results might not be accurate, and the federation would be vulnerable to threats while collecting and updating model parameters. Joint sharing modeling of multi-party data is made possible by the use of blockchain technology, which is integrated with federated learning to meet specific data-sharing requirements. Table \ref{table:2} provides the existing blockchain-enabled FL methodologies designed and developed by several researchers. These methods can be used in financial services for improving the service quality. Because the blockchain has a strong way to store certificates, it is possible to provide controllable data traceability. For federated learning, both making data visible and making it available are possible. As a result, the functions that they perform are somewhat complementary to one another. If they can be connected, it will not only be possible to improve the way data is shared, but it will also be possible to keep data modeling private \cite{28}.
\par \noindent \textbf{Data Protection} 
Feng and Chen \cite{29} proposed a blockchain-enabled FL data-preserving strategy, which can be used by a financial business. Each business node that joins the consortium chain must pass a certain qualification test. Business nodes that take part in federated learning within the consortium chain must also verify their strong reputation value, but only if it is higher than a certain threshold. In federated learning, only nodes with a certain minimum level of reputation can be chosen to take part. They divided federated learning nodes into monitoring and training groups. The training node group does local model training rounds. The monitoring node group determines the training node reputation and global parameters. The top m reputation-valued nodes in this cycle of federated learning are the monitoring nodes. This is done to make it even less likely that malicious nodes will be included in the process of learning.
\par \noindent \textbf{Data Traceability}Federated computing ensures that the original data never leaves the local node, that only the gradient information of the model update needs to be uploaded, and that private data is protected while taking advantage of the blockchain's immutability, certificate storage function strength, and controllable traceability. Based on this concept, it is possible to train complicated joint models. Therefore, they serve a similar purpose but in different ways. When used in tandem, they may not only make data exchange a reality but also keep personal information secure.

\begin{table*}[!hbp]
\scriptsize
\begin{center}
\caption{\textbf{Existing Blockchain-enabled FL Methods for Data sharing}}\label{table:2}
\begin{tabular}{ |m{3em}|m{13.5em}|m{15.3em}|m{16.3em} |m{16.3em}|} 
\hline
\textbf{Author / Ref} &\textbf{Proposed Method} &  \textbf{Description} &  \textbf{Pros} & \textbf{Cons} \\
\hline 

{Feng \textit{et al.,}\newline \cite{29}} & {Credibility-based RAFT efficient consensus method}
 & {Used Consortium blockchain and federated learning to enterprise data sharing and monitor the process.} &{$\bullet$High enterprise application efficiency.
\newline $\bullet$Enterprise data exchange without compromising privacy.} &{$\bullet$It is unable to defend itself against Byzantine attacks}
 \\
\hline

{Zhao \textit{et al.,}\newline \cite{30}} & {Reputation-based FL method}
 & {This strategy trains a machine-learning model using consumer data. Manufacturers can then estimate client needs and expenditures. } 
 
 &{$\bullet$ Introduced a reward system.
\newline $\bullet$Entice clients to the crowdsourced FL activity.
\newline $\bullet$Provide privacy and test accuracy.} 
&{$\bullet$Deterministically optimum local-global balance for greater test accuracy.}
 \\
\hline

{Lu \textit{et al.,}\newline \cite{31}} & {Privacy-preserving distributed data sharing} & {This strategy trains a machine-learning model using consumer data. Manufacturers can then estimate client needs and expenditures. } 
 &{$\bullet$ Improved computational resource efficiency.
\newline $\bullet$Improved data exchange efficiency.
\newline $\bullet$Secure information exchange with higher accuracy.} 
&{$\bullet$Data utility.
\newline $\bullet$Inefficient data sharing method, if the number of devices is limited}
\\
\hline

{Bandara \textit{et al.,}\newline \cite{32}} & {Bassa-ML}  & {An integrated federated learning platform that is built on blockchain technology and Model Cards. } 
 &{$\bullet$More openness and transparency
\newline $\bullet$Bringing audit ability to the process of federated learning.
\newline $\bullet$Enhanced trust for the models
} 
 &{$\bullet$ Data utility

}
\\
\hline

{Li \textit{et al.,}\newline \cite{33}} & {BFLC}  & { Blockchain is used by the framework in place of a centralized server for both the storing of global model data and the communication of local model update information.} 
 &{$\bullet$Cut down on the amount of computing needed for consensus. 
\newline $\bullet$Reduce the number of harmful assaults
} 
 &{$\bullet$ Scalability of the BFLC.
\newline $\bullet$Storage is Expensive because blockchain stores all info.
}
\\
\hline

{Kim \textit{et al.,}\newline \cite{34}} & {BlockFL}  & {Architecture for distributed learning that allows for the secure and transparent sharing and verification of localized modifications to learning models.} 
 &{$\bullet$Incentive-based method
\newline $\bullet$Federates additional devices with more training samples.
} 
 &{$\bullet$ Uncontrolled data quality
}
\\
\hline

{Ramanan \textit{et al.,}\newline \cite{35}} & {BAFFLE}  & { Decomposing the global parameter space into discrete pieces and then using a score and bid technique is how BAFFLE increases computing efficiency.} 
 &{$\bullet$ Update many sections simultaneously
\newline $\bullet$ A low cost for the computation.

} 
 &{$\bullet$No method for data quality control
}
\\
\hline

{Shayan \textit{et al.,}\newline \cite{36}} & {Biscotti}  & {A completely decentralized peer-to-peer (P2P) method for multi-party machine learning, as well as a client-to-client machine learning procedure that protects users' privacy. } 
 &{$\bullet$ The PoF consensus algorithm was introduced.
\newline $\bullet$Scalable, resistant to errors, and able to withstand recognized forms of assault
} 
 &{$\bullet$ Scalability issues for the large model.
\newline $\bullet$Privacy issues.
\newline $\bullet$Stake limitations.
}
\\
\hline

{ Chen \textit{et al.,}\newline \cite{37}} & {LearningChain}  & { Decentralized privacy-preserving and secure machine learning system using a broad (linear or nonlinear) deep learning approach with no centralized trusted servers} 
 &{ $\bullet$A decentralized framework
\newline $\bullet$Protection against attacks performed by Byzantine
} 
 &{$\bullet$ Prone to being influenced by biases
}
\\
\hline

{Lyu \textit{et al.,}\newline \cite{38}} & {FPPDL}  & {A local credibility mutual assessment method for fairness and a three-layer onion-style encryption technique for accuracy and privacy. } 
 &{$\bullet$ Privacy-preserved
\newline $\bullet$Fairness
\newline $\bullet$Accuracy
} 
 &{$\bullet$ - Many encryption methods may impede node processing.
\newline $\bullet$Fairness in Non-IID setting.
}
\\
\hline
{Ma \textit{et al.,}\newline \cite{39}} & {BLADE-FL}  & {Prevent dishonest customers from tainting the learning process, and further give clients a learning environment that is self-motivating and dependable. } 
 &{$\bullet$ Privacy preserved
\newline $\bullet$Lazy client protection
} 
 &{$\bullet$ - Computing capability.
\newline $\bullet$Training data size
\newline $\bullet$Transmitting diversity.
}
\\
\hline
\end{tabular}
\end{center}
\end{table*}

\par \noindent \textbf{Profit Sharing} Because nodes in a community can use the model whenever they want without making any changes, a strong incentive is needed to get nodes to train models. Li et al. \cite{33} suggested that a system called ``profit sharing by contribution" be put in place as a way to solve this problem. Following the aggregation of each round, the managers will then give incentives to the appropriate nodes based on the scores of the updates that they have provided. As a result, giving updates regularly may bring more benefits, and the constantly updated global model may encourage other nodes to join the network. It is important to research this incentive mechanism since it can easily be adapted to a variety of applications in the real world and has great scalability.   
\par \noindent \textbf{Storage Optimization} When it comes to applications that take place in the real world, storage overhead is a critical component that helps determine the hardware requirements for the training devices. Historical models and updates may help with recovery after a disaster, but they take up a huge amount of space. Here, Li et al. \cite{33} gave a simple and workable plan for reducing the storage overhead: nodes that don't have enough space can get rid of old blocks locally, and they only need to keep the most recent model and updates for the current round. In this way, the problem of not having enough storage space can be solved, and the core nodes can keep their ability to recover and verify data. Having said that, it is also easy to see where this approach falls short. With each transaction that is deleted, the legitimacy of the blockchain suffers. It is possible that each node will not adopt this technique due to the high level of mutual mistrust that exists among this group. So, the best thing to do might be to store your data with a reputable and trustworthy third party. The only information that the blockchain stores are the network addresses of the locations of each model or updated file, as well as records of any modifications that were made to those files. For other nodes to access the model and its changes, the centralized store must interface with those nodes. This central storage will be in charge of making backups in case of a disaster and offering services for storing files in different places. 
\par \noindent \textbf{Safeguarding Payment Networks}
One of the best things about blockchain technology is that it can be used as a payment network that is not limited by national borders. As a decentralized answer to the problem of making payments without friction and with low transaction costs, many different blockchain protocols have been made. Centralized financial institutions are known for their high fees and painfully slow processing times, which led to the creation of these alternatives. Even though this use of blockchain technology has a lot of potentials, it is still hard to use it widely because of security concerns \cite{77}. Theft and fraud are all too frequent in blockchain transactions since all that is required to complete them is a set of public and private keys. On the other hand, advanced machine learning can easily find unusual account activity, which then calls for human help. Both the companies that supply financial services and the people who use them are protected by this extra security measure \cite{78}.
\par \noindent \textbf{Effective Financial Services}
Financial institutions that use FL and blockchain technologies often have the goal of increasing both the speed at which they provide their services and the quality of those services. In the same way that any other company would, these establishments have the incentive to find ways to save expenses and, as a result, generate greater value. In a survey that shows this trend, Deloitte found \cite{40} that 57\% of firms see cost savings as the main benefit of joining consortium blockchain networks. Those that supply financial services may deliver more value to their consumers while also optimizing the returns on their investments if they use these two technologies to drive business operations.

\par \noindent \textbf{Controlled Finance Automation} The trend toward more automation cannot be denied, but if it is allowed to continue unchecked, it may result in undesirable consequences. Businesses will always lose control of their operations over time if there are no limits on how they can use automated processes. As a result of this, the tasks of automation need to be carried out in conjunction with built-in checks and balances. Using FL and smart contracts backed by blockchain technology could help make this dynamic happen. Smart contacts make it possible to automate procedures, while machine learning can look for problems and only call for human help when it's necessary. Because of this very important infrastructure, all financial transactions would be completely safe, completely open, and completely efficient.

\section{Future Scopes}
Financial companies have always been concerned about cybersecurity breaches, but cyber-attacks now pose considerably greater hazards to their operations and reputations. They must know the biggest threats first. It will help them prioritize cybersecurity activities and maximize ROI. It will also aid cybersecurity strategy development. Ransomware, phishing, online application, and vulnerability exploitation assaults, denial of service (DoS) attacks \cite{41}, insider threats \cite{42}, nation-state attacks, and state-sponsored threats will be the biggest dangers to financial institutions in 2022. Financial institutions must upgrade their security to protect against ever-changing threats. Despite spending millions and adding several levels of protection to their infrastructure, they still don't know how to employ security measures. To remain ahead of threat actors, test their IT security infrastructure against real-world threats often. IT and security executives must verify security measures to assess firms' cybersecurity posture and cyber resilience and show that they prevent intrusions. 

\par \noindent \textbf{Effectiveness of Transmission}
The storage and synchronization of the blockchain need a significant amount of hardware resources, including not only the space on the hard drive but also the bandwidth on the network. Therefore, the best way to lessen the amount of transmission that is needed while still preserving the reliability of model training is a subject deserving of further investigation.

\par \noindent \textbf{The Scene in Public}
The authentication processes are handled by the alliance blockchain system, although this does have the side effect of making it more difficult to join the training community. How to create a public community by using a Proof-of-Work System Another interesting problem is finding a way to work together while protecting yourself from attacks from hostile nodes in financial services.

\par \noindent \textbf{Lightweight Training}
A lot of the Internet of Things (IoT) devices that financial clients use don't have enough hardware features to train a deep neural network well. Because of this, the question of how to make training models easier (for example, by using edge servers) while still protecting users' privacy is an important one that deserves more research.

\begin{figure}[!t]
\centering
\includegraphics[width=0.5\textwidth]{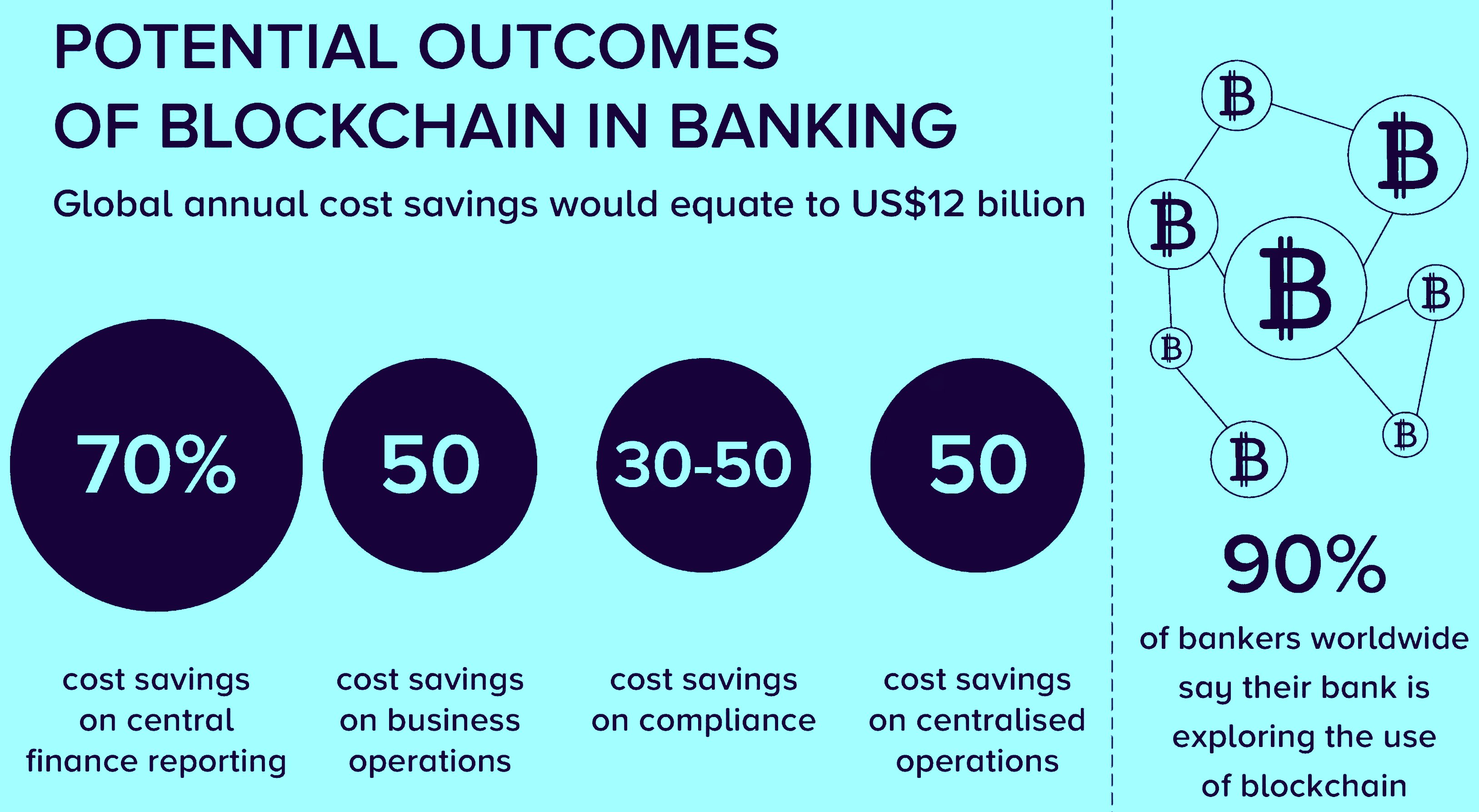}
 \caption{The potential outcomes of using blockchain technology in the financial sector. \cite{3}}
\label{fig6}
\vspace{-0.5cm} 
\end{figure}

\par \noindent \textbf{Security Control Validation}
Even though it's good to see financial institutions work toward high levels of cyber maturity, we strongly suggest doing an objective evaluation of these assumptions and maturity levels and fixing any gaps between what was expected and what was found in an assessment. Banks and other financial institutions, as well as other businesses, must constantly test the efficacy of their security policies against real-world attacks to stay one step ahead of threat actors. The term ``security control validation" refers to an approach that is centered on potential threats and that enables businesses to evaluate and analyze their cybersecurity posture and their overall cyber resilience. It also checks to see if the security controls are working well enough to stop cyberattacks.
\par \noindent \textbf{Automation}
As long as AI keeps getting good investments, FL seems to be in a great position to grow a lot. Because of the adaptability of the technology, it is expected that it will make its way into a rising number of different businesses across a wide variety of use cases. The financial services industry, in particular, is in a great position to get a lot of value out of the combination of blockchain and FL technologies. These technologies work together to make big changes in the financial sector by making it safer, making it run better, and giving people more control over automation. Fig. \ref{fig6} shows the potential outcomes of using blockchain technology in the financial sector.

\par \noindent \textbf{Next-Generation Marketing}
However, to make use of this potential, it is essential to keep in mind why blockchain technology was developed in the first place. Also, for FL learning systems to keep being useful, they need to be fed with good enough data. Although many questions remain unanswered, these technologies are not going away, either together or separately. In a financial market that is always changing, blockchain and FL integrations might be the next drivers of disruptive change across the finance sector.

\section{Conclusions}
FL, a distributed AI technique, is attracting attention for its privacy-enhancing and scalable financial services and applications. We conducted a state-of-the-art assessment and comprehensive talks based on recent research to examine how blockchain and FL may improve financial services. This study was motivated by the lack of a comprehensive FL and blockchain survey in financial services. We first discussed FL and blockchain technologies and their combination to bridge this gap. Then, we extended our study by providing measures of how FL and blockchain can be used in financial services to protect financial data, make decentralized storage more efficient, keep payment networks safe, automate tasks, protect privacy, and keep things safe. Lastly, we talked about some of the problems with research and gave notions for future initiatives that will bring more attention to this new field and encourage more research to realize FL and blockchain.

\bibliographystyle{IEEEtran}
\bibliography{Fin.bib}

\begin{thebibliography}{10}
\providecommand{\url}[1]{#1}
\csname url@samestyle\endcsname
\providecommand{\newblock}{\relax}
\providecommand{\bibinfo}[2]{#2}
\providecommand{\BIBentrySTDinterwordspacing}{\spaceskip=0pt\relax}
\providecommand{\BIBentryALTinterwordstretchfactor}{4}
\providecommand{\BIBentryALTinterwordspacing}{\spaceskip=\fontdimen2\font plus
\BIBentryALTinterwordstretchfactor\fontdimen3\font minus
  \fontdimen4\font\relax}
\providecommand{\BIBforeignlanguage}[2]{{%
\expandafter\ifx\csname l@#1\endcsname\relax
\typeout{** WARNING: IEEEtran.bst: No hyphenation pattern has been}%
\typeout{** loaded for the language `#1'. Using the pattern for}%
\typeout{** the default language instead.}%
\else
\language=\csname l@#1\endcsname
\fi
#2}}
\providecommand{\BIBdecl}{\relax}
\BIBdecl

\bibitem{1}
Y.~Zhang, Z.~Wang, J.~Deng, Z.~Gong, I.~Flood, and Y.~Wang, ``Framework for a
  blockchain-based infrastructure project financing system,'' \emph{IEEE
  Access}, vol.~9, pp. 141\,555--141\,570, 2021.

\bibitem{56}
K.~Mantzaris, ``How can we benefit from blockchain technologies?''
  \url{https://economistmk.blogspot.com/2018/03/how-can-we-benefit-from-blockchain.html},
  2018, [Online; Accessed on Jan. 18, 2023].

\bibitem{2}
Chirag, ``Blockchain in financial services: A catalyst for disruption in
  finance world,'' \url{https://appinventiv.com/blog/blockchain-and-fintech/},
  2022, [Online; Accessed on Jan. 13, 2023].

\bibitem{4}
J.~Nicholls, A.~Kuppa, and N.-A. Le-Khac, ``Financial cybercrime: A
  comprehensive survey of deep learning approaches to tackle the evolving
  financial crime landscape,'' \emph{IEEE Access}, vol.~9, pp.
  163\,965--163\,986, 2021.

\bibitem{5}
D.~Das, S.~Banerjee, P.~Chatterjee, U.~Ghosh, and U.~Biswas, ``A secure
  blockchain enabled v2v communication system using smart contracts,''
  \emph{IEEE Transactions on Intelligent Transportation Systems}, pp. 1--10,
  2022.

\bibitem{6}
\BIBentryALTinterwordspacing
D.~Das, K.~Dasgupta, and U.~Biswas, ``A secure blockchain-enabled vehicle
  identity management framework for intelligent transportation systems,''
  \emph{Computers and Electrical Engineering}, vol. 105, p. 108535, 2023.
  [Online]. Available:
  \url{https://www.sciencedirect.com/science/article/pii/S0045790622007509}
\BIBentrySTDinterwordspacing

\bibitem{7}
L.~Bhatia and S.~Samet, ``Decentralized federated learning: A comprehensive
  survey and a new blockchain-based data evaluation scheme,'' in \emph{2022
  Fourth International Conference on Blockchain Computing and Applications
  (BCCA)}, 2022, pp. 289--296.

\bibitem{3}
Chirag, ``Why are banks adopting blockchain technology?''
  \url{https://appinventiv.com/blog/blockchain-in-banking/}, 2023, [Online;
  Accessed on Jan. 18, 2023].

\bibitem{8}
Polestar, ``Top financial services banking analytics use cases,''
  \url{https://www.polestarllp.com/top-financial-services-banking-analytics-use-cases},
  2022, [Online; Accessed on Jan. 10, 2023].

\bibitem{9}
D.~Donaldson, ``Vulnerability of financial institutions to cyber crime,''
  \url{https://www.iap-association.org/getattachment/Conferences/Regional-Conferences/North-America-and-Caribbean/4th-North-American-and-Caribbean-Conference/Conference-Documentation/4NACC_Jamaica_WS2B_PPT_Damien_Donaldson.pdf.aspx},
  2022, [Online; Accessed on Dec. 18, 2022].

\bibitem{10}
Swivelsecure, ``5 cybersecurity weaknesses in banking and finance,''
  \url{https://swivelsecure.com/solutions/banking-finance/5-cybersecurity-weaknesses-threats-in-banking-and-finance-industry/},
  2022, [Online; Accessed on Dec. 12, 2022].

\bibitem{11}
T.~Maurer and A.~Nelson, ``The global cyber threat,” finance \&
  development,''
  \url{https://www.imf.org/external/pubs/ft/fandd/2021/03/pdf/global-cyber-threat-to-financial-systems-maurer.pdf},
  2021, [Online; Accessed on Sep. 11, 2022].

\bibitem{12}
E.~Kost, ``The 6 biggest cyber threats for financial services in 2023,''
  \url{https://www.upguard.com/blog/biggest-cyber-threats-for-financial-services},
  2023, [Online; Accessed on Jan. 15, 2023].

\bibitem{13}
R.~Gall, ``Over 600,000 sites impacted by wp statistics patch,''
  \url{https://www.wordfence.com/blog/2021/05/over-600000-sites-impacted-by-wp-statistics-patch/},
  [Online; Accessed on Aug. 22, 2022].

\bibitem{15}
Adminer, ``Database management in a single php file,''
  \url{https://www.adminer.org/en/ }, 2021, [Online; Accessed on Nov. 17,
  2022].

\bibitem{16}
A.~Community, ``Confluence server and data center - cve-2021-26084, confluence
  server webwork ognl injection,''
  \url{https://confluence.atlassian.com/doc/confluence-security-advisory-2021-08-25-1077906215.html},
  2021, [Online; Accessed on Dec. 17, 2022].

\bibitem{17}
Intellipaat, ``The importance of cyber security in banking sector: Cyber
  security in banking,''
  \url{https://intellipaat.com/blog/cyber-security-in-banking/}, 2022, [Online;
  Accessed on Sep. 05, 2022].

\bibitem{18}
F.~C. Feeney, ``Cybersecurity regulation harmonization: The financial services
  roundtable – bits,''
  \url{https://www.hsgac.senate.gov/imo/media/doc/Testimony-Feeney-2017-06-21.pdf},
  2017, [Online; Accessed on Nov. 25, 2022].

\bibitem{19}
P.~Schroeder, ``Capital one to pay \$80 million fine after data breach,''
  \url{https://www.reuters.com/article/us-usa-banks-capital-one-fin-idUSKCN2522DA},
  2020, [Online; Accessed on Sep. 06, 2022].

\bibitem{20}
A.~Bronstad, ``Capital one reaches \$190m settlement over 2019 data breach,''
  \url{https://www.law.com/2021/12/21/capital-one-settles-lawsuits-over-2019-data-breach/},
  2021, [Online; Accessed on May 19, 2022].

\bibitem{21}
S.~Özarslan, ``Six stages of dealing with a global security incident,''
  \url{https://www.picussecurity.com/resource/blog/six-stages-of-dealing-with-a-global-security-incident},
  2021, [Online; Accessed on May 27, 2022].

\bibitem{22}
``Csbs 2022 national survey of community banks, findings from the 2022 csbs
  national survey of community banks presented at the 10th annual community
  banking research conference,''
  \url{https://www.picussecurity.com/resource/blog/six-stages-of-dealing-with-a-global-security-incident},
  2022, [Online; Accessed on Oct. 24, 2022].

\bibitem{57}
O.~Evans, ``Blockchain technology and the financial market: an empirical
  analysis,'' 2018.

\bibitem{59}
L.~Cocco, A.~Pinna, and M.~Marchesi, ``Banking on blockchain: Costs savings
  thanks to the blockchain technology,'' \emph{Future internet}, vol.~9, no.~3,
  p.~25, 2017.

\bibitem{58}
T.~Goud~Allam, A.~B.~M. Mehedi~Hasan, A.~Maag, and P.~Prasad, ``Ledger
  technology of blockchain and its impact on operational performance of banks:
  A review,'' in \emph{2021 6th International Conference on Innovative
  Technology in Intelligent System and Industrial Applications (CITISIA)},
  2021, pp. 1--10.

\bibitem{60}
V.~Br{\"u}hl, ``Virtual currencies, distributed ledgers and the future of
  financial services,'' \emph{Intereconomics}, vol.~52, no.~6, pp. 370--378,
  2017.

\bibitem{43}
J.~Liu, L.~Yan, and D.~Wang, ``A hybrid blockchain model for trusted data of
  supply chain finance,'' \emph{Wireless Personal Communications}, vol. 127,
  pp. 919 -- 943, 2021.

\bibitem{44}
J.~Lorenz, M.~I. Silva, D.~O. Apar{\'i}cio, J.~T. Ascens{\~a}o, and P.~Bizarro,
  ``Machine learning methods to detect money laundering in the bitcoin
  blockchain in the presence of label scarcity,'' \emph{Proceedings of the
  First ACM International Conference on AI in Finance}, 2020.

\bibitem{45}
Y.-H. Chen, L.-C. Huang, I.-C. Lin, and M.-S. Hwang, ``Research on the secure
  financial surveillance blockchain systems.'' \emph{Int. J. Netw. Secur.},
  vol.~22, no.~4, pp. 708--716, 2020.

\bibitem{46}
S.~Chen, Q.~Wang, and S.~Liu, ``Credit risk prediction in peer-to-peer lending
  with ensemble learning framework,'' in \emph{2019 Chinese Control And
  Decision Conference (CCDC)}, 2019, pp. 4373--4377.

\bibitem{47}
\BIBentryALTinterwordspacing
M.~Kherbouche, G.~Pisoni, and B.~Molnár, ``Model to program and blockchain
  approaches for business processes and workflows in finance,'' \emph{Applied
  System Innovation}, vol.~5, no.~1, 2022. [Online]. Available:
  \url{https://www.mdpi.com/2571-5577/5/1/10}
\BIBentrySTDinterwordspacing

\bibitem{48}
C.~Ma, X.~Kong, Q.~Lan, and Z.~Zhou, ``The privacy protection mechanism of
  hyperledger fabric and its application in supply chain finance,''
  \emph{Cybersecurity}, vol.~2, no.~1, pp. 1--9, 2019.

\bibitem{49}
M.~Du, Q.~Chen, J.~Xiao, H.~Yang, and X.~Ma, ``Supply chain finance innovation
  using blockchain,'' \emph{IEEE Transactions on Engineering Management},
  vol.~67, no.~4, pp. 1045--1058, 2020.

\bibitem{50}
B.~Wu and T.~Duan, ``Application blockchain in supply chain finance: a study on
  small and micro enterprises in xi'an,'' in \emph{2021 2nd International
  Conference on Big Data Economy and Information Management (BDEIM)}, 2021, pp.
  479--482.

\bibitem{51}
R.~D. Camino, R.~State, L.~Montero, and P.~Valtchev, ``Finding suspicious
  activities in financial transactions and distributed ledgers,'' in \emph{2017
  IEEE International Conference on Data Mining Workshops (ICDMW)}, 2017, pp.
  787--796.

\bibitem{52}
Y.~Guo and C.~Liang, ``Blockchain application and outlook in the banking
  industry,'' \emph{Financial innovation}, vol.~2, no.~1, pp. 1--12, 2016.

\bibitem{53}
J.~Chod, N.~Trichakis, G.~Tsoukalas, H.~Aspegren, and M.~Weber, ``On the
  financing benefits of supply chain transparency and blockchain adoption,''
  \emph{Management Science}, vol.~66, no.~10, pp. 4378--4396, 2020.

\bibitem{54}
D.~Li, D.~Han, N.~Crespi, R.~Minerva, and K.-C. Li, ``A blockchain-based secure
  storage and access control scheme for supply chain finance,'' \emph{The
  Journal of Supercomputing}, pp. 1--30, 2022.

\bibitem{55}
D.~Li, D.~Han, and H.~Liu, ``Fabric-chain \& chain: a blockchain-based
  electronic document system for supply chain finance,'' in \emph{International
  Conference on Blockchain and Trustworthy Systems}.\hskip 1em plus 0.5em minus
  0.4em\relax Springer, 2020, pp. 601--608.

\bibitem{23}
M.~Guerar, A.~Merlo, M.~Migliardi, F.~Palmieri, and L.~Verderame, ``A
  fraud-resilient blockchain-based solution for invoice financing,'' \emph{IEEE
  Transactions on Engineering Management}, vol.~67, no.~4, pp. 1086--1098,
  2020.

\bibitem{74}
W.-T. Tsai, R.~Blower, Y.~Zhu, and L.~Yu, ``A system view of financial
  blockchains,'' in \emph{2016 IEEE Symposium on Service-Oriented System
  Engineering (SOSE)}.\hskip 1em plus 0.5em minus 0.4em\relax IEEE, 2016, pp.
  450--457.

\bibitem{65}
X.~Han, Y.~Yuan, and F.-Y. Wang, ``A blockchain-based framework for central
  bank digital currency,'' in \emph{2019 IEEE International Conference on
  Service Operations and Logistics, and Informatics (SOLI)}.\hskip 1em plus
  0.5em minus 0.4em\relax IEEE, 2019, pp. 263--268.

\bibitem{66}
C.~Barontini and H.~Holden, ``Proceeding with caution-a survey on central bank
  digital currency,'' \emph{Proceeding with Caution-A Survey on Central Bank
  Digital Currency (January 8, 2019). BIS Paper}, no. 101, 2019.

\bibitem{67}
D.~Andolfatto, ``Assessing the impact of central bank digital currency on
  private banks,'' \emph{The Economic Journal}, vol. 131, no. 634, pp.
  525--540, 2021.

\bibitem{64}
H.~Hassani, X.~Huang, E.~S. Silva, H.~Hassani, X.~Huang, and E.~S. Silva,
  \emph{Fusing Big Data, blockchain, and cryptocurrency}.\hskip 1em plus 0.5em
  minus 0.4em\relax Springer, 2019.

\bibitem{24}
M.~Du, Q.~Chen, J.~Xiao, H.~Yang, and X.~Ma, ``Supply chain finance innovation
  using blockchain,'' \emph{IEEE Transactions on Engineering Management},
  vol.~67, no.~4, pp. 1045--1058, 2020.

\bibitem{62}
N.~Dashkevich, S.~Counsell, and G.~Destefanis, ``Blockchain application for
  central banks: A systematic mapping study,'' \emph{IEEE Access}, vol.~8, pp.
  139\,918--139\,952, 2020.

\bibitem{63}
F.~Holotiuk, F.~Pisani, and J.~Moormann, ``The impact of blockchain technology
  on business models in the payments industry,'' 2017.

\bibitem{68}
P.~K. Ozili, ``Blockchain finance: Questions regulators ask,'' in
  \emph{Disruptive innovation in business and finance in the digital
  world}.\hskip 1em plus 0.5em minus 0.4em\relax Emerald Publishing Limited,
  2019.

\bibitem{69}
Q.~K. Nguyen, ``Blockchain-a financial technology for future sustainable
  development,'' in \emph{2016 3rd International conference on green technology
  and sustainable development (GTSD)}.\hskip 1em plus 0.5em minus 0.4em\relax
  IEEE, 2016, pp. 51--54.

\bibitem{70}
R.~Auer, ``Embedded supervision: how to build regulation into blockchain
  finance,'' 2019.

\bibitem{75}
X.~Wang, X.~Xu, L.~Feagan, S.~Huang, L.~Jiao, and W.~Zhao, ``Inter-bank payment
  system on enterprise blockchain platform,'' in \emph{2018 IEEE 11th
  international conference on cloud computing (CLOUD)}.\hskip 1em plus 0.5em
  minus 0.4em\relax IEEE, 2018, pp. 614--621.

\bibitem{76}
T.~Wu and X.~Liang, ``Exploration and practice of inter-bank application based
  on blockchain,'' in \emph{2017 12th international conference on computer
  science and education (ICCSE)}.\hskip 1em plus 0.5em minus 0.4em\relax IEEE,
  2017, pp. 219--224.

\bibitem{71}
A.~Hayes, ``Decentralized banking: Monetary technocracy in the digital age,''
  in \emph{Banking beyond banks and money}.\hskip 1em plus 0.5em minus
  0.4em\relax Springer, 2016, pp. 121--131.

\bibitem{72}
J.~Caytas, ``Developing blockchain real-time clearing and settlement in the eu,
  us, and globally,'' \emph{Columbia Journal of European Law: Preliminary
  Reference (June 22, 2016)}, 2016.

\bibitem{73}
R.~Priem, ``Distributed ledger technology for securities clearing and
  settlement: benefits, risks, and regulatory implications,'' \emph{Financial
  Innovation}, vol.~6, no.~1, pp. 1--25, 2020.

\bibitem{25}
W.~Zheng, Z.~Zheng, X.~Chen, K.~Dai, P.~Li, and R.~Chen, ``Nutbaas: A
  blockchain-as-a-service platform,'' \emph{IEEE Access}, vol.~7, pp.
  134\,422--134\,433, 2019.

\bibitem{26}
T.-M. Choi, ``Financing product development projects in the blockchain era:
  Initial coin offerings versus traditional bank loans,'' \emph{IEEE
  Transactions on Engineering Management}, vol.~69, no.~6, pp. 3184--3196,
  2022.

\bibitem{27}
D.~Das, S.~Banerjee, U.~Ghosh, U.~Biswas, and A.~K. Bashir, ``A decentralized
  vehicle anti-theft system using blockchain and smart contracts,''
  \emph{Peer-to-Peer Networking and Applications}, vol.~14, p. 2775–2788,
  2021.

\bibitem{61}
Y.-H. Chen, L.-C. Huang, I.-C. Lin, and M.-S. Hwang, ``Research on the secure
  financial surveillance blockchain systems.'' \emph{Int. J. Netw. Secur.},
  vol.~22, no.~4, pp. 708--716, 2020.

\bibitem{28}
D.~Das, S.~Banerjee, and U.~Biswas, ``A secure vehicle theft detection
  framework using blockchain and smart contract,'' \emph{Peer-to-Peer
  Networking and Applications}, vol.~14, pp. 672--686, 2021.

\bibitem{29}
X.~Feng and L.~Chen, ``Data privacy protection sharing strategy based on
  consortium blockchain and federated learning,'' in \emph{2022 International
  Conference on Artificial Intelligence and Computer Information Technology
  (AICIT)}, 2022, pp. 1--4.

\bibitem{30}
Y.~Zhao, J.~Zhao, L.~Jiang, R.~Tan, D.~Niyato, Z.~Li, L.~Lyu, and Y.~Liu,
  ``Privacy-preserving blockchain-based federated learning for iot devices,''
  \emph{IEEE Internet of Things Journal}, vol.~8, no.~3, pp. 1817--1829, 2021.

\bibitem{31}
Y.~Lu, X.~Huang, Y.~Dai, S.~Maharjan, and Y.~Zhang, ``Blockchain and federated
  learning for privacy-preserved data sharing in industrial iot,'' \emph{IEEE
  Transactions on Industrial Informatics}, vol.~16, no.~6, pp. 4177--4186,
  2020.

\bibitem{32}
E.~Bandara, S.~Shetty, A.~Rahman, R.~Mukkamala, J.~Zhao, and X.~Liang,
  ``Bassa-ml — a blockchain and model card integrated federated learning
  provenance platform,'' in \emph{2022 IEEE 19th Annual Consumer Communications
  \& Networking Conference (CCNC)}, 2022, pp. 753--759.

\bibitem{33}
Y.~Li, C.~Chen, N.~Liu, H.~Huang, Z.~Zheng, and Q.~Yan, ``A blockchain-based
  decentralized federated learning framework with committee consensus,''
  \emph{IEEE Network}, vol.~35, no.~1, pp. 234--241, 2021.

\bibitem{34}
H.~Kim, J.~Park, M.~Bennis, and S.-L. Kim, ``Blockchained on-device federated
  learning,'' \emph{IEEE Communications Letters}, vol.~24, no.~6, pp.
  1279--1283, 2020.

\bibitem{35}
P.~Ramanan and K.~Nakayama, ``Baffle : Blockchain based aggregator free
  federated learning,'' in \emph{2020 IEEE International Conference on
  Blockchain (Blockchain)}, 2020, pp. 72--81.

\bibitem{36}
M.~Shayan, C.~Fung, C.~J.~M. Yoon, and I.~Beschastnikh, ``Biscotti: A
  blockchain system for private and secure federated learning,'' \emph{IEEE
  Transactions on Parallel and Distributed Systems}, vol.~32, no.~7, pp.
  1513--1525, 2021.

\bibitem{37}
X.~Chen, J.~Ji, C.~Luo, W.~Liao, and P.~Li, ``When machine learning meets
  blockchain: A decentralized, privacy-preserving and secure design,'' in
  \emph{2018 IEEE International Conference on Big Data (Big Data)}, 2018, pp.
  1178--1187.

\bibitem{38}
L.~Lyu, J.~Yu, K.~Nandakumar, Y.~Li, X.~Ma, and J.~Jin, ``Towards fair and
  decentralized privacy-preserving deep learning with blockchain,''
  \emph{ArXiv}, vol. abs/1906.01167, 2019.

\bibitem{39}
C.~Ma, J.~Li, L.~Shi, M.~Ding, T.~Wang, Z.~Han, and H.~V. Poor, ``When
  federated learning meets blockchain: A new distributed learning paradigm,''
  \emph{IEEE Computational Intelligence Magazine}, vol.~17, no.~3, pp. 26--33,
  2022.

\bibitem{77}
W.-T. Tsai, E.~Deng, X.~Ding, and J.~Li, ``Application of blockchain to trade
  clearing,'' in \emph{2018 ieee international conference on software quality,
  reliability and security companion (qrs-c)}.\hskip 1em plus 0.5em minus
  0.4em\relax IEEE, 2018, pp. 154--163.

\bibitem{78}
M.~C. Okoye and J.~Clark, ``Toward cryptocurrency lending,'' in \emph{Financial
  Cryptography and Data Security: FC 2018 International Workshops, BITCOIN,
  VOTING, and WTSC, Nieuwpoort, Cura{\c{c}}ao, March 2, 2018, Revised Selected
  Papers 22}.\hskip 1em plus 0.5em minus 0.4em\relax Springer, 2019, pp.
  367--380.

\bibitem{40}
Z.~Chen, ``What the convergence of blockchain and machine learning means for
  the future of finance,''
  \url{https://www.nasdaq.com/articles/what-the-convergence-of-blockchain-and-machine-learning-means-for-the-future-of-finance},
  2023, [Online; Accessed on Jan. 15, 2023].

\bibitem{41}
C.~Kemp, C.~Calvert, and T.~M. Khoshgoftaar, ``Detecting slow application-layer
  dos attacks with pca,'' in \emph{2021 IEEE 22nd International Conference on
  Information Reuse and Integration for Data Science (IRI)}, 2021, pp.
  176--183.

\bibitem{42}
S.~Chaipa, E.~K. Ngassam, and S.~Shawren, ``Towards a new taxonomy of insider
  threats,'' in \emph{2022 IST-Africa Conference (IST-Africa)}, 2022, pp.
  1--10.

\end{thebibliography}

\end{document}